\documentclass[fleqn,usenatbib]{mnras}

\usepackage[T1]{fontenc}
\usepackage{ae,aecompl}

\usepackage{graphicx}	
\usepackage{savesym}
\usepackage{amsmath}
\savesymbol{iint}
\savesymbol{iiint}
\usepackage{txfonts}
\restoresymbol{TXF}{iint}
\restoresymbol{TXF}{iiint}
\usepackage{amssymb}	
\usepackage{hyperref}
\usepackage{url}
\usepackage{pdflscape}
\usepackage{microtype}
\usepackage{color}
\usepackage[utf8]{inputenc}
\usepackage{comment}
\usepackage{subfigure}

\newcommand{\nh}{$n_{\rm H}$}
\newcommand{\z}{$Z$}
\newcommand{\dnmax}{$\Delta_{\rm max} (\log n_{\rm H})$}
\newcommand{\dzmax}{$\Delta_{\rm max} (\log Z)$}
\newcommand{\dmax}{$\Delta_{\rm max}$}

\defcitealias{Khaire15ebl}{KS15b}
\defcitealias{Khaire15puc}{KS15a}
\defcitealias{KS19}{KS19}
\defcitealias{HM12}{HM12}
\defcitealias{Madau15}{MH15}
\defcitealias{FG09}{FG09}
\defcitealias{FG20}{FG20}
\defcitealias{Puchwein19}{P19}

\graphicspath{{./}{figures/}}

\title[CGM \& UV background]{\huge How Robust are the Inferred Density and Metallicity of the Circumgalactic Medium? 
}

\author[Acharya \& Khaire]
{
\parbox{\textwidth}{
Anshuman Acharya$^{1,2}$\thanks{E-mail: anshuman@mpa-garching.mpg.de} and Vikram Khaire$^{3,4}$ 
} 
\vspace*{10pt}\\
$^{1}$Indian Institute of Science Education and Research (IISER) - Mohali, SAS Nagar, Punjab - 140306, INDIA\\
$^{2}$Max Planck Institut für Astrophysik, Garching - 85748, Germany \\
$^{3}$Indian Institute of Space Science \& Technology, Thiruvananthapuram, Kerala -  695547, INDIA\\
$^{4}$Physics Department, Broida Hall, University of California Santa Barbara, Santa Barbara, CA 93106-9530, USA\\
}   
\date{}
\pubyear{2020}

\begin{document}

\begin{sloppypar}
\maketitle
\label{firstpage}
\pagerange{\pageref{firstpage}--\pageref{lastpage}} 
\begin{abstract}
Quantitative estimates of the basic properties of the circumgalactic medium 
(CGM), such as its density and metallicity, depend on the spectrum of incident 
UV background radiation. 
Models of UV background are known to have large variations, mainly because
they are synthesized using poorly constrained parameters, which
introduce uncertainty in the inferred properties of the CGM. 
Here, we quantify this uncertainty 
using a large set of new UV background models with physically
motivated toy models of metal-enriched CGM. 
We find that, the inferred density and metallicity of low-density 
($10^{-5}$ cm$^{-3}$) gas is uncertain by factors of 6.3 and 3.2, whereas
high density ($10^{-3}$ cm$^{-3}$) gas by factors of 4 and 1.6, respectively. 
The variation in the shape of the UV background models
is entirely responsible for such a variation in the metallicity 
while variation in the density arises from both normalization and shape
of the UV background. Moreover, we find a harder (softer) UV background
infers higher (lower) density and metallicity. 
We also study warm-hot gas at $T= 10^{5.5}$ K and find that 
metallicity is robustly estimated but the inferred density is 
uncertain by a factor of 3 to 5.4 for low to high-density gas. 
Such large uncertainties in density and metallicity may severely limit the 
studies of the CGM and demand better
observational constraints on the input parameters used in synthesizing UV background. 
\end{abstract}


\section{Introduction} \label{sec:intro}

Galaxies form, feed, and grow by accreting pristine gas from the 
intergalactic medium (IGM) and spew metal-enriched gas into vast distances. 
Much of this active trade of gas determines the evolution of galaxies. 
The region where this trade happens is located around galaxies at the interface
between their discs and the IGM, known as the circumgalactic medium (CGM).
Therefore, studying the gas in the CGM, its distribution, chemical 
composition and physical conditions around different types of galaxies may 
play a crucial role in understanding the galaxy formation and 
evolution \citep[see review;][]{Tumlinson17}. 

In recent times, much of the insights on the CGM around different galaxies 
have been obtained from promising large observational campaigns 
\citep[e.g.,][]{Stoke13_cgm, Tumlinson13_cos_halos, Borodoloi14_cos_dwarf, 
Burchett19_casbah, Chen20_cubs} and dedicated 
efforts to simulate the small scale structure of the CGM 
\citep[e.g.,][]{Hummels13_enzo_cgm, Hummels19_tempest, Churchill15_sim_cgm,Dylan20_tng_cgm,
Oppenheimer20_eagle_cgm, 
Freeke19_cgm_sim, Peeples19_foggie}. 
There is an important ingredient needed for both, simulating the CGM as well as 
gaining insights from observations, known as the 
ionizing UV background radiation. 
For the absorption-line studies of CGM absorbers, 
this all-prevailing spectrum of the 
UV background is needed for determining 
the physical conditions and chemical composition of the CGM gas.  
This is achieved by relating the observed ionic abundances to 
the total abundances of gas by performing an ionization correction. 
The results of which determine the density and metallicity of gas
which can be used to quantify the mass and line-of-sight 
size of the CGM absorbers and their total metal content 
\citep[e.g.,][]{Stoke13, Shull14Metals, Prochaska17, Mohapatra19, Mohapatra21}. 
Moreover, the spectrum of the UV background is 
essential to determine if the absorbing gas is single-phase or multi-phase 
\citep[e.g.,][]{Tripp08, Narayanan09, Savage14, Sameer21},
photoionized or collisionally ionized 
\citep[e.g.,][]{Hussain15, Hussain17, Pachat17, Haislmaier21}, 
or if it shows any enhancement in the relative abundances 
of the metals compared to the solar 
abundances \citep[e.g.,][]{Pettini08, Cooke11, Shull14Metals}.
Mostly, individual CGM studies assume a single model of the UV background 
in their analysis and it is not always feasible, or generally expected, 
to use a range of all possible UV background models. 

\begin{figure}

\includegraphics[width=0.46\textwidth,height=\textheight,keepaspectratio]{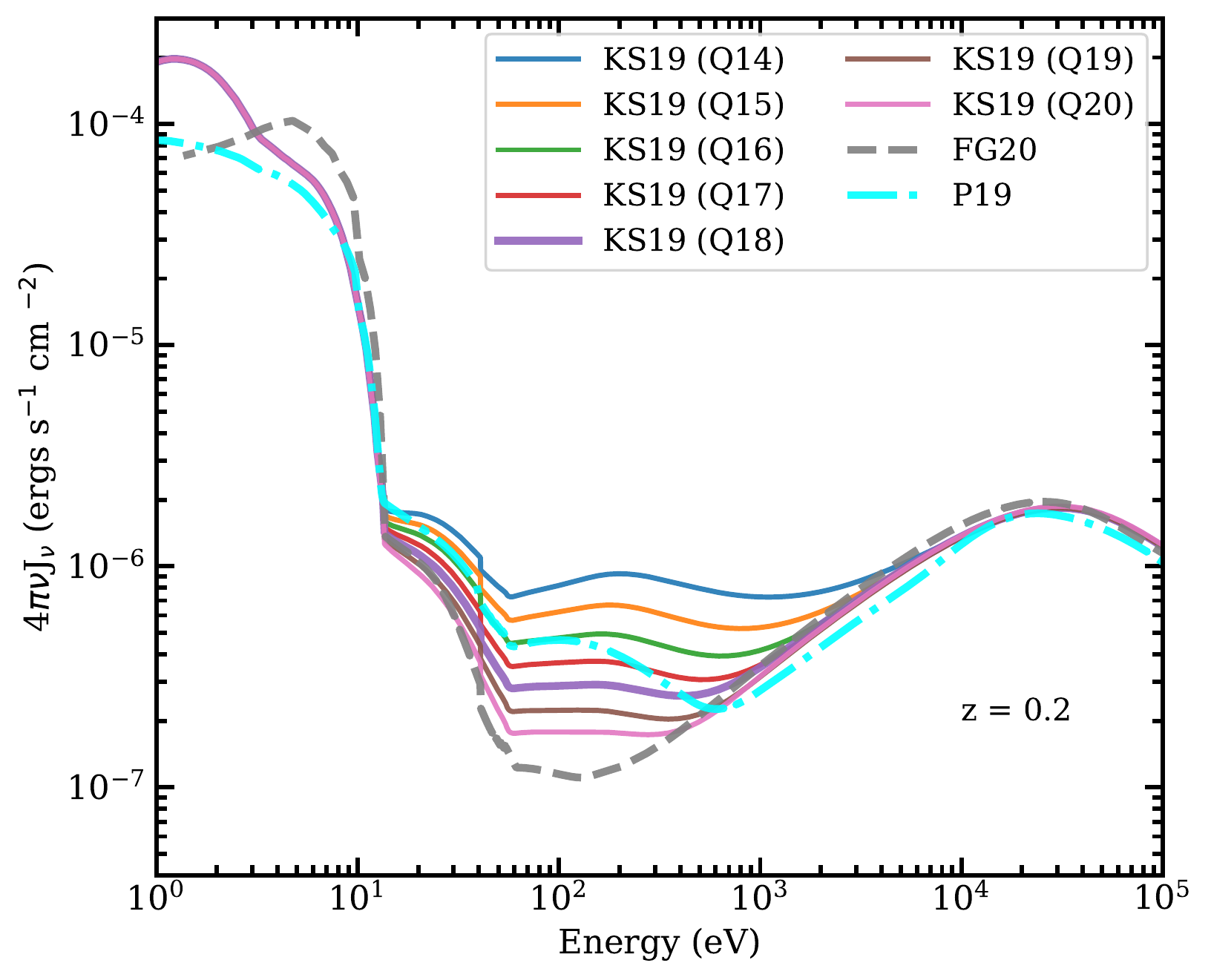}
\caption{
The UV background spectrum at $z= 0.2$. Solid lines show \citetalias{KS19} 
UV background models obtained by varying intrinsic quasar SED 
(a power-law $L_{\nu} \propto \nu ^{-\alpha}$) index $\alpha$ 
from 1.4 (model Q14) to 2.0 (model Q20). We also show two new UV 
background models by \citetalias{Puchwein19} and \citetalias{FG20}. 
Even though all of these models show large variation, they are
the most updated versions of the UV background and all of them are 
physically viable (see Fig.~\ref{fig.gamma_HI}). 
}
\label{fig.uvb}
\end{figure}

Because direct observations of the UV background are not possible, 
it needs to be synthesized via a cosmological 
radiative transfer of UV 
photons emitted by quasars and galaxies through the IGM
\citep[see][]{Miralda90, HM96, Shull99}. 
Therefore, the spectrum of the synthesized 
UV background depends on the UV emissivity of quasars and galaxies,
the distribution of gas in the IGM and also on various assumptions that are part
of the modeling \citep[for more details see section 4.4 of][]{KS19}. 
In particular, it is more susceptible to the
poorly constrained emissivities of quasars and galaxies 
and shows large variation at any redshift even
among the latest models (see Fig.~\ref{fig.uvb}) obtained 
using the most updated
observations of inputs \citep[][]{KS19, Puchwein19, FG20}. 
Such a variation in the UV background models can lead to a varying degree
of inferred properties of the CGM absorbers and reported results depend
on the choice of UV background models. This was illustrated previously 
by \citet{Hussain17} for the case of Ne~{\sc viii} absorbers and
\citet{Chen17} for metallicity estimates. 
Given the large uncertainty in the 
UV background models, it is important to know how much variation (or a systematic
uncertainty) one expects in the inferred basic properties of the CGM gas, 
such as its density and metallicity. Past research using older UV background models like \citet{HM12} and \citet{FG09} have noted systematic uncertainties in metallicity like in \citet{Lehner_2013} and more recently \citet{Wotta_2019} for cool, photoionized gas.
In this paper, we quantify these variations with the help of widely used {\sc cloudy} software \citep[][]{Ferland98, Ferland17} 
for ionization modelling and a set of nine latest UV background models. 
We use seven viable UV background models as described in \citet{KS19} (hereafter 
\citetalias{KS19}) where the spectral energy distribution (SED) of quasars was varied to obtain the 
quasar emissivity in the UV background models. Such a variation
is within the reported observational range and also consistent with high-$z$ 
observations of He~{\sc ii} \citep[see][]{Khaire17sed}.

In addition to these seven models,  
we also use two more latest UV background models; \citet{Puchwein19} 
(hereafter \citetalias{Puchwein19}) and 
\citet{FG20} (hereafter \citetalias{FG20}), which are updated versions of
previous UV background models by \citet{HM12} and \citet{FG09}, respectively. 
For illustrating the differences, we show all these nine UV background 
models at $z =0.2$ in Fig.~\ref{fig.uvb} and discuss the details in Section~\ref{sec:2}.

Using {\sc cloudy} \citep[v17.02;][]{Ferland17}, 
we irradiate a gas cloud  with  a preferred UV background model
and create mock CGM observations. We use two types of {\sc cloudy} 
models for creating 
mock observations, in first type of models,
the absorber is kept in thermal and 
photoionization equilibrium with 
the UV background, while in second type the absorber is 
a warm-hot gas having high temperatures ($T = 10^{5.5}$ K). 
We note down the output column densities of many metal ions from 
these models and treat a subset of those as our mock CGM observations. 
These mock observations are then used
to infer the density and metallicity of the absorbing gas with
different UV background models. 
Difference between the inferred values of density and metallicity
for a UV background model and
the true values (that are used for generating mock observations with a preferred
UV background), quantifies the expected variation arising from using different
UV background models.  

We find that, by varying the UV background models
the inferred density and metallicity of low-density 
($10^{-5}$ cm$^{-3}$) gas can be off by factors of 6.3 and 3.2, whereas
high density ($10^{-3}$ cm$^{-3}$) gas can be off by
factors of 4 and 1.6, respectively. 
For warm-hot absorber at $T= 10^{5.5}$ K, we find that 
metallicity is robustly estimated but the inferred density can be off
by a factor of 3 to 5.4 for low to high-density gas. 
In addition to this, we also discuss the relative effect of shape and
normalization of UV background models on the inferred densities and metallicities.

The paper is organized as follows. In Section \ref{sec:2}, we briefly discuss the
UV background models used in the study.
In Section \ref{sec:3}, we explain our mock CGM observations and 
present the method to infer the
density and metallicity of the CGM gas from those. 
In section \ref{sec:res} we present our results for the different UV background models,
and quantify the variation in the inferred density 
and metallicity for both photoionized and warm-hot absorbers. 
In section \ref{sec:5} we summarize main results of the paper.

\begin{figure}
\includegraphics[width=0.49\textwidth,height=\textheight,keepaspectratio]{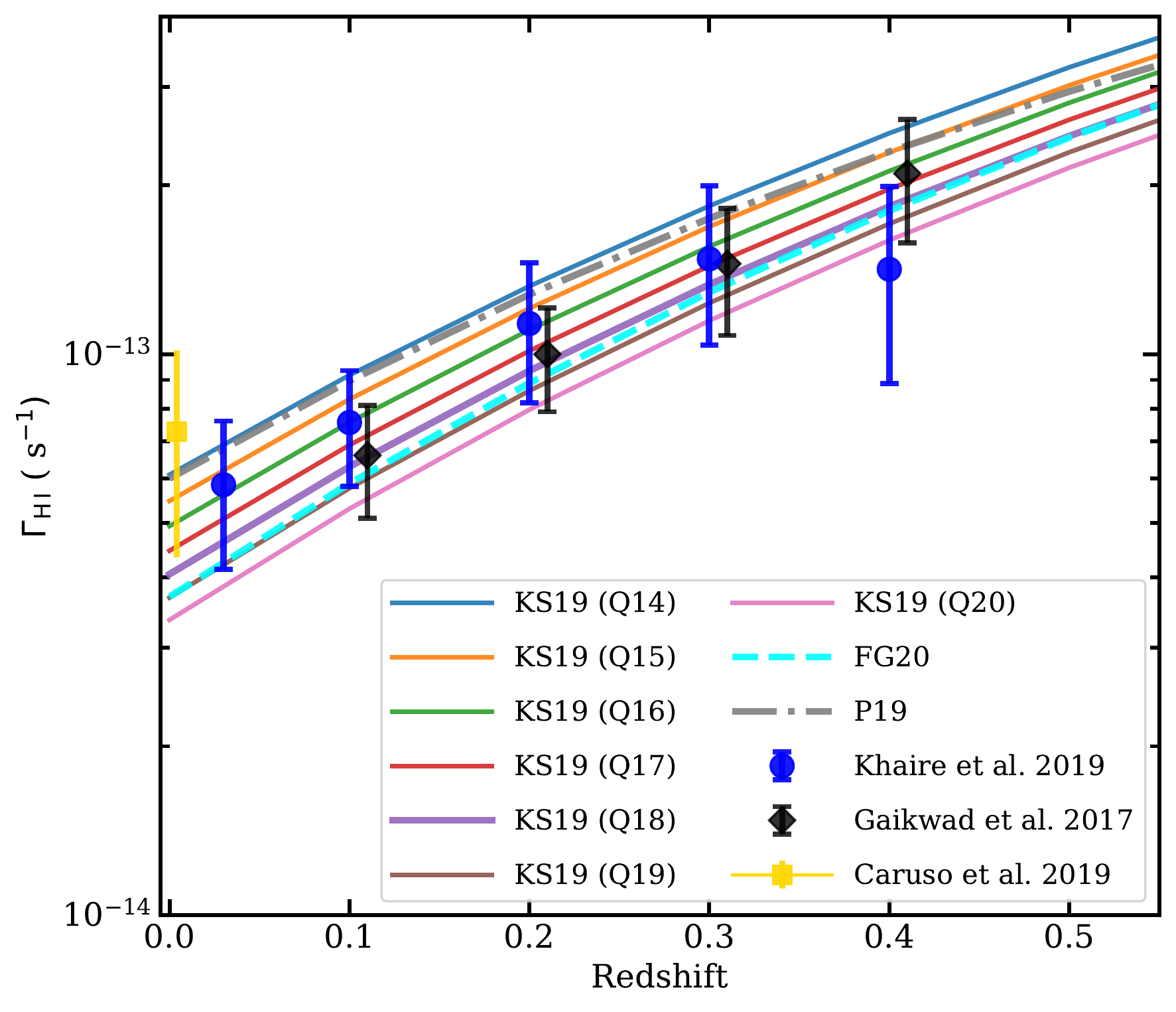}
\caption{
The H~{\sc i} photoionization rates ($\Gamma_{\rm HI}$)
with redshift ($z<0.5$) for different UV background models along with the
measurements \citep[][]{Gaikwad17a, Khaire19ps, Caruso19}. 
Solid lines show \citetalias{KS19} UV background models obtained by varying 
intrinsic quasar SED (a power-law $L_{\nu} \propto \nu ^{-\alpha}$) 
index $\alpha$ from 1.4 (model Q14) to 2.0 (model Q20). All these models,including 
two new UV background models by \citetalias{Puchwein19} and 
\citetalias{FG20}, are consistent with the $\Gamma_{\rm HI}$ measurements.
Moreover, they encompass the whole $1-\sigma$ range in the 
$\Gamma_{\rm HI}$ measurements.}
\label{fig.gamma_HI}
\end{figure}

\section{The UV background models} \label{sec:2}

We use seven UV background models from \citetalias{KS19} and two recent 
additional models from
\citetalias{Puchwein19} and \citetalias{FG20}. 
In Fig.\ref{fig.uvb}, we show the predictions of these nine 
models at $z = 0.2$. There is a large variation in the model 
intensity of UV background at energy range $10-10^3$ eV, 
notably almost an order of
magnitude around $50-500$ eV. 
Ionization potentials of most of the metal ions that are
typically observed in the CGM fall in the energy range $\sim 10-250$ eV.
Therefore, a large variation in the UV background spectrum among different 
models, especially at these energies,  can have significant
impact on the inferred properties of the CGM gas.

Most of the variation in the UV background models, as seen in 
Fig.\ref{fig.uvb}, can be accounted by the difference 
in the emissivity of sources used as input. 
UV background models assume two types of sources; galaxies and quasars. 
At low redshifts $z \lesssim 2$, there is now consensus that the UV background 
(at $E > 13.6$ eV) is contributed by quasars alone 
\citep[][]{Khaire15puc, Gaikwad17a, Kulkarni18, Khaire19ps}, and hence
the contribution from galaxies is negligibly small. 
Therefore, the variation in the UV background models seen at low-$z$ 
(see Fig.~\ref{fig.uvb}) is primarily due to 
differences in the quasar emissivity. There are other factors such as the
H~{\sc i} column density distribution used in the UV background models and
also the model prediction of He {\sc ii} column densities
\citep[see][]{Khaire13}, however their contribution to the variation 
in the UV background spectrum 
is subdominant as compared to the quasar emissivity at $z<2$. 

The main reason for differences in the quasar emissivity at $10-10^3$ eV is 
the lack of good constraints on the mean quasar SED. Quasar emissivity is 
modelled by using a power-law 
$L_{\nu} \propto \nu ^{-\alpha}$ at $E \gtrsim 10$ eV
whose reported measurements do not agree with each other and scatter in their 
values imprints large differences on the emissivity and hence on 
the UV background models. Even the most recent studies with
large data-sets \citep[][]{Shull12sed, Stevans14, Tilton16, Lusso15, Lusso18} 
report values of power-law index $\alpha$ ranging from $0.7$ to $2.2$ 
for mean quasar SED. Moreover, the highest 
energy at which the power-law SEDs are measured is less than $30$ eV. 
Therefore, in absence of any constraints on the quasar SED for $E>30$ eV, UV 
background models use the same power-law at $E> 13.6$ eV and
extrapolate  it till soft X-rays around $1-2$ keV 
\citepalias[][]{KS19, Puchwein19}. 
One exception is the new model by \citetalias{FG20}, where author 
connected the SED at $E= 21$ eV to 1 keV which 
resulted into softer emissivity and therefore softer 
UV background at higher energies
as compared to other models (see Fig.~\ref{fig.uvb}). 
Given the huge range in reported values of power-law index $\alpha$ for quasar 
SED, \citetalias{KS19} provided seven UV background models
by varying $\alpha$ from $1.4$ (model Q14) to $2.0$ (model Q20) with an increment of 0.1. This small conservative range of $\alpha = 1.4-2$ as compared to the reported observations by different studies (i.e $\alpha = 0.7-2.2$) was chosen to be consistent with high-$z$ observations of He~{\sc ii}
\citep[$\alpha  = 1.6-2.0$ from][]{Khaire17sed} and to incorporate new low-$z$ measurements of $\alpha$ by \citet{Shull12sed} and \citet{Stevans14}. 
All of these seven models published in \citepalias[][]{KS19} at $z=0.2$ are shown in 
Fig. \ref{fig.uvb}.  Every other parameters in these models are same and only the
soft X-ray SED (at $E \gtrsim 1$ keV) has been modified for each model 
to match the X-ray background measurements (for more details see \citetalias[][]{KS19}).
Whereas, both \citetalias{Puchwein19} and 
\citetalias{FG20} chose $\alpha = 1.7$ but latter used it 
till $E \sim 21$ eV only. Even though the quasar SED is same in three models 
Q17 of \citetalias{KS19}, \citetalias{Puchwein19} and \citetalias{FG20}, the UV 
background spectra are different because of differences in the
input parameters such as quasar emissivity at $13.6$ eV, 
its redshift evolution and the H~{\sc i} distribution in the IGM. 
As compared to the models of \citetalias{Puchwein19} and \citetalias{FG20}, 
one of the advantage
of using seven models of \citetalias{KS19} is that 
they differ only in terms of quasar SED 
(at $13 {\rm eV} \lesssim E \lesssim 1 {\rm keV}$ determined by $\alpha$). 
Note that we are not synthesizing any new UV background models 
in this paper,
rather we are just using existing models. For more details on individual models 
we direct readers to respective literature  \citepalias[i.e][]{KS19, FG20, Puchwein19}. 

All the nine UV background 
models are physically viable options. This is illustrated in Fig.~\ref{fig.gamma_HI}
where we show the predicted H~{\sc i} photoionization rates ($\Gamma_{\rm HI}$) from 
all nine UV background models along with $\Gamma_{\rm HI}$ measurements at $z<0.5$.
Measurement of $\Gamma_{\rm HI}$ are shown from \citet{Khaire19ps} and 
\citet{Gaikwad17a} that are obtained using low-$z$ Lyman-$\alpha$ forest 
\citep[from][]{Danforth14} while \citet{Caruso19} measurement
used H-$\alpha$ fluorescence from outskirts of nearby galaxy UGC 7321.
Despite different shapes, all nine UV background models
are consistent with $\Gamma_{\rm HI}$ measurements. 
Moreover, these models approximately cover the range in $\Gamma_{\rm HI}$ measurements
and their uncertainty. Therefore,
these models are ideal set of models to find the 
systematic variation in the inferred CGM properties. 
Ideally, for CGM observations 
one should find the systematic uncertainty in the inferred quantities 
by using all these models, and quote it along with statistical
uncertainty of the inferred quantities but it is not always feasible for individual 
CGM studies to do that.
Therefore, in this paper we quantify the average systematic uncertainty on the
inferred density and metallicity arising from UV background variation. For that we used 
toy CGM observations and two inference techniques as explained in the next section.
Also note that there are still many studies that
choose old UV background models such as \citet{HM12} and \citet{FG09} that 
are shown to be inconsistent with new observations 
\citep[see][]{Kollmeier14, Shull15, Khaire15puc, Gaikwad17a, Gaikwad17b}, 
which can add more to the systematic uncertainty in the inferred properties of the CGM.

\section{Toy CGM observations and methods to infer the density and metallicity}\label{sec:3}
In this section, we discuss our toy models that serve as mock CGM observations
for photoionized and warm-hot gas. We also discuss two different methods to 
infer the hydrogen number density $n_{\rm H}$ (cm$^{-3}$)
and metallicity $Z/Z_{\odot}$ from those mock observations.

\subsection{Toy models of photoionized and warm-hot absorbers}\label{sec.toy_mode}

\begin{figure*}
\begin{tabular}{cc}
  \includegraphics[width=85mm]{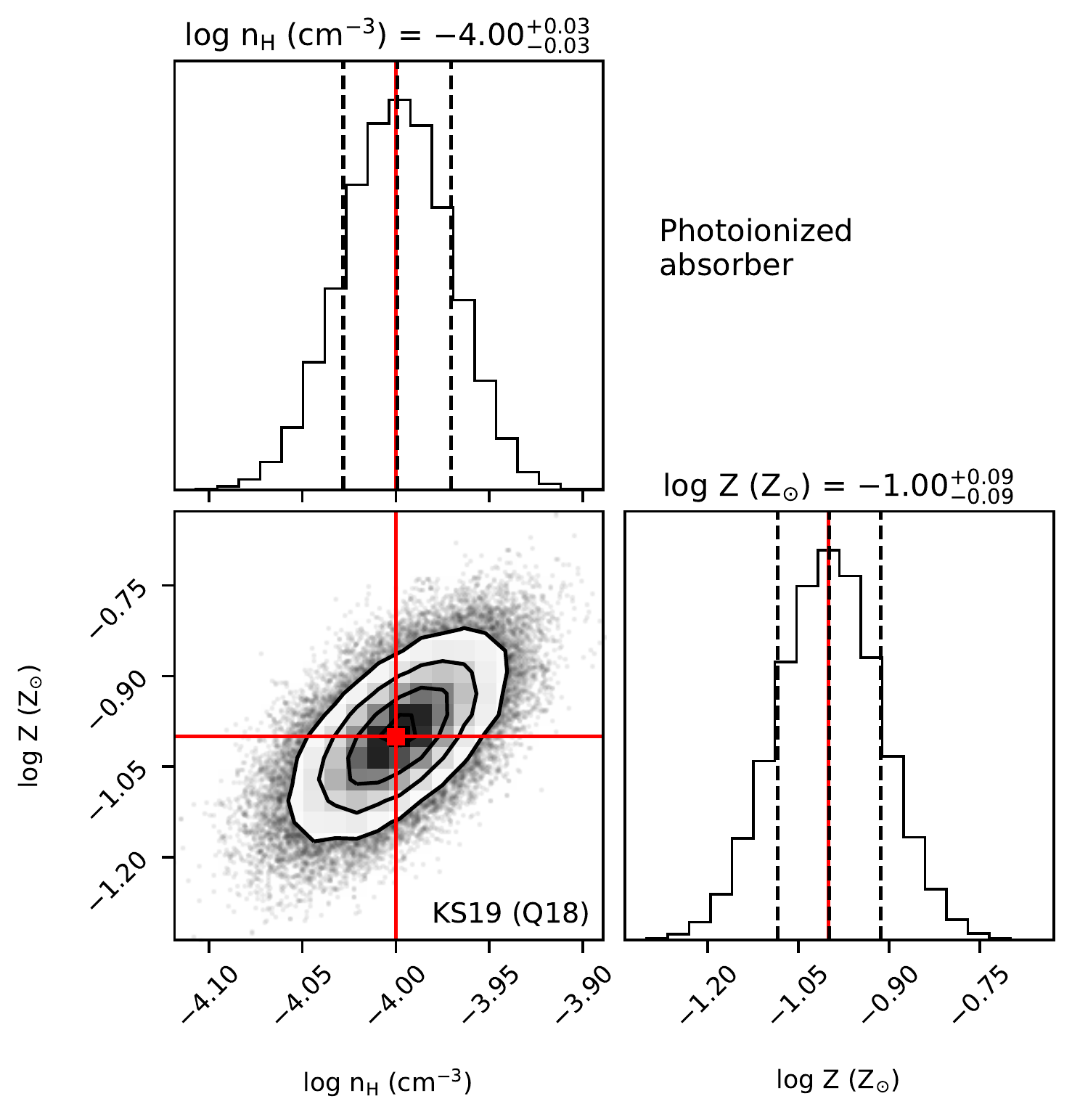} &   \includegraphics[width=85mm]{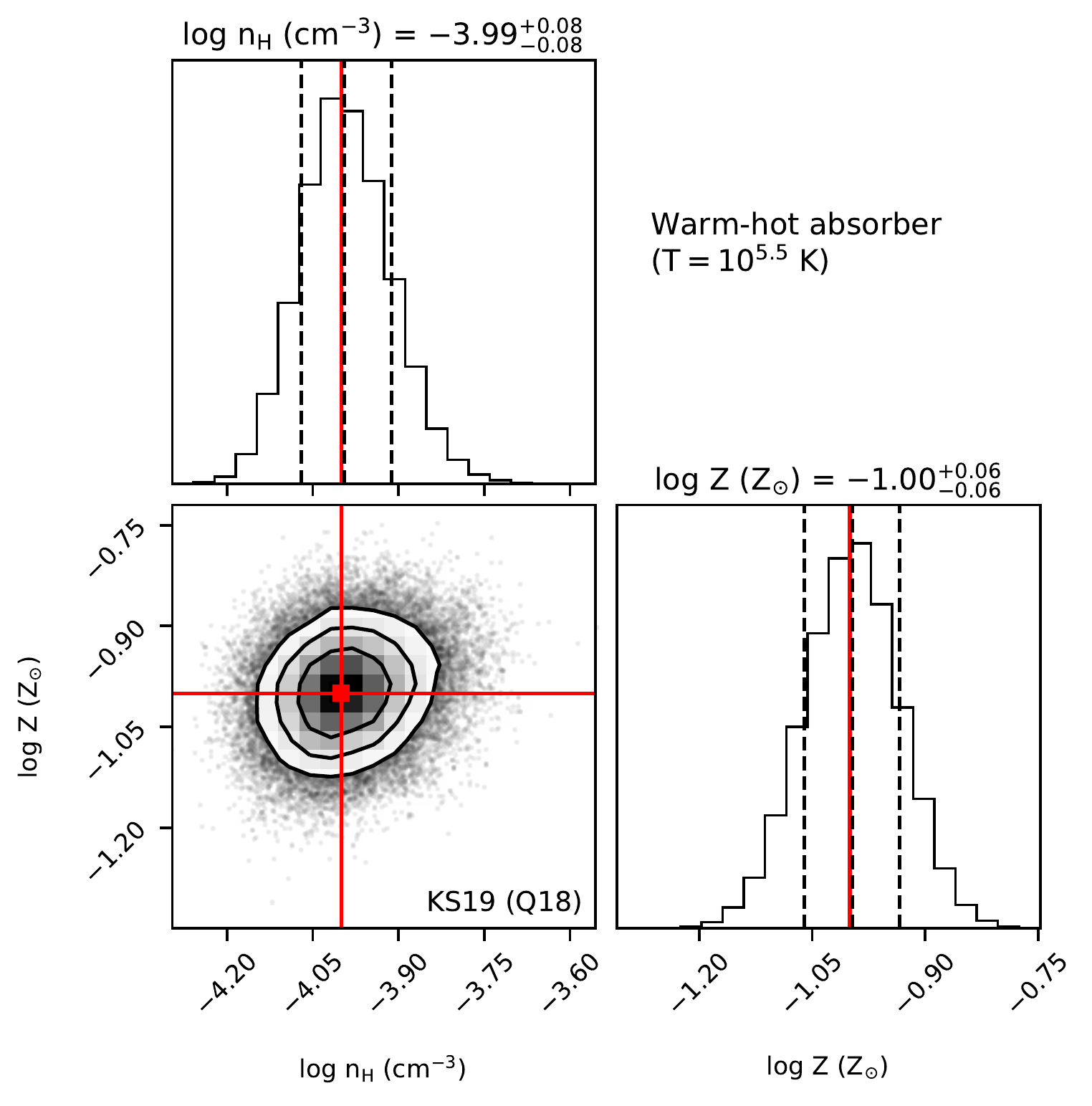} \\
\end{tabular}
\caption{A joint posterior PDF of $\log n_{\rm H }$ (cm $^{-3}$) and $\log Z$ 
(Z$_{\odot}$) along with their 
marginalized PDF are shown for the inference test for photoionized absorber 
(left-hand panel) and warm-hot absorber at $T= 10^{5.5}$ K (right-hand panel). 
We use the same Q18 
\citetalias{KS19} UV background to infer the $n_{\rm H }$ and $Z$ 
that is used in our toy model to create mock CGM observations.
Red lines shows true values of $\log n_{\rm H} (\rm cm^{-3}) = -4$ and 
$\log Z (Z_{\rm \odot}) = -1$ which are in excellent agreement with the median of
marginalized PDFs as shown in the legends (with errors indicating a range of 16 and 84 percentiles).}
\label{fig.inference}
\end{figure*}

{\sc cloudy} ionization models are routinely used for inferring
hydrogen number density $n_{\rm H}$ (cm$^{-3}$)
and metallicity $Z/Z_{\odot}$ of CGM absorbers. In such models, 
these absorbers are assumed to represent plane parallel 
slabs of gas with uniform density and metallicity in a thermal and 
photoionization equilibrium with the extragalactic UV background.
The basic methodology is to vary the density and metallicity in {\sc cloudy} 
models to obtain a good agreement between model column densities and observed column 
densities of all ion-species in an absorber. 
Taking cue from this, we set up a plane parallel slab absorber in {\sc cloudy v17.02} 
which is in ionization and thermal equilibrium with 
Q18 model of \citetalias{KS19} (their fiducial model with $\alpha = 1.8$)

having hydrogen number density $n_{\rm H} = 10 ^{-4}$ cm$^{-3}$ and
metallicity $Z = 0.1 Z_{\odot}$. 
This absorber is assumed to have neutral hydrogen column
density $N_{\rm H I} = 10^{15}$ cm$^{-2}$, which is used as a stopping 
criteria for the {\sc cloudy} model. From the output of this {\sc cloudy} 
model we note-down column densities of various metal ions. 
A set of column densities $\{\log N^{\rm obs}_{\rm ion}\}$ from those
along with $N_{\rm H I} = 10^{15}$ cm$^{-2}$ constitute our toy
CGM observation. We call it as our `photoionized' mock absorber.
Later, we change these input values used for modelling mock absorber to provide general results.

Given the fact that a significant amount of gas in the IGM and CGM resides in a
warm-hot $10^5 - 10^7$ K phase \citep[see][]{Shull12baryons}, we also generate 
another mock observations for such a gas. 
For that, we keep the plane parallel slab of gas at a constant
temperature $T= 10^{5.5}$ K and irradiated it with same  Q18 \citetalias{KS19} 
UV background in {\sc cloudy}. All the other parameters in these {\sc cloudy}
models are same as the 
`photoionized' mock absorber described above 
(i.e $n_{\rm H} = 10 ^{-4}$ cm$^{-3}$, $Z = 0.1 Z_{\rm \odot}$ and $N_{\rm HI} = 10 ^{15}$ cm$^{-2}$).
Similar to the photoionized case, we note down the column densities of various 
metal ions from {\sc cloudy} output, and use a small 
$\{\log N^{\rm obs}_{\rm ion}\}$ subset out of it
along with $N_{\rm H I} = 10^{15}$ cm$^{-2}$ 
as our mock observation for the warm-hot gas. 
We call it as a `warm-hot' mock absorber.
In all {\sc cloudy} models (ran for mock observations as well as for the inference), 
we take solar abundance of heavy elements from \citet{Grevesse10}. 
We also scale the elemental helium abundance by a factor $8.163 \times 10^{-2}$ 
so that the ratio of hydrogen to helium number density 
($n_{\rm H}/n_{\rm He}$) becomes 
$12.25$, consistent with recent measurements of helium mass fraction 
$Y_{p} = 0.246^{+0.039}_{-0.041}$ (where $n_{\rm H}/ n_{\rm He} = 4(1-Y_p)/Y_p$) from cosmic
microwave background \citep[][with TT+ LowE]{Planck18}.
Without such a scaling the default elemental abundance of helium in {\sc cloudy} 
represents the one typically found it the interstellar medium 
i.e $n_{\rm H}/n_{\rm He} = 10$.

\subsection{Methods to jointly constrain $n_{\rm H}$ and $Z$ }\label{sec:3.2}
\begin{figure*}
\begin{tabular}{cc}
  \includegraphics[width=0.82\textwidth]{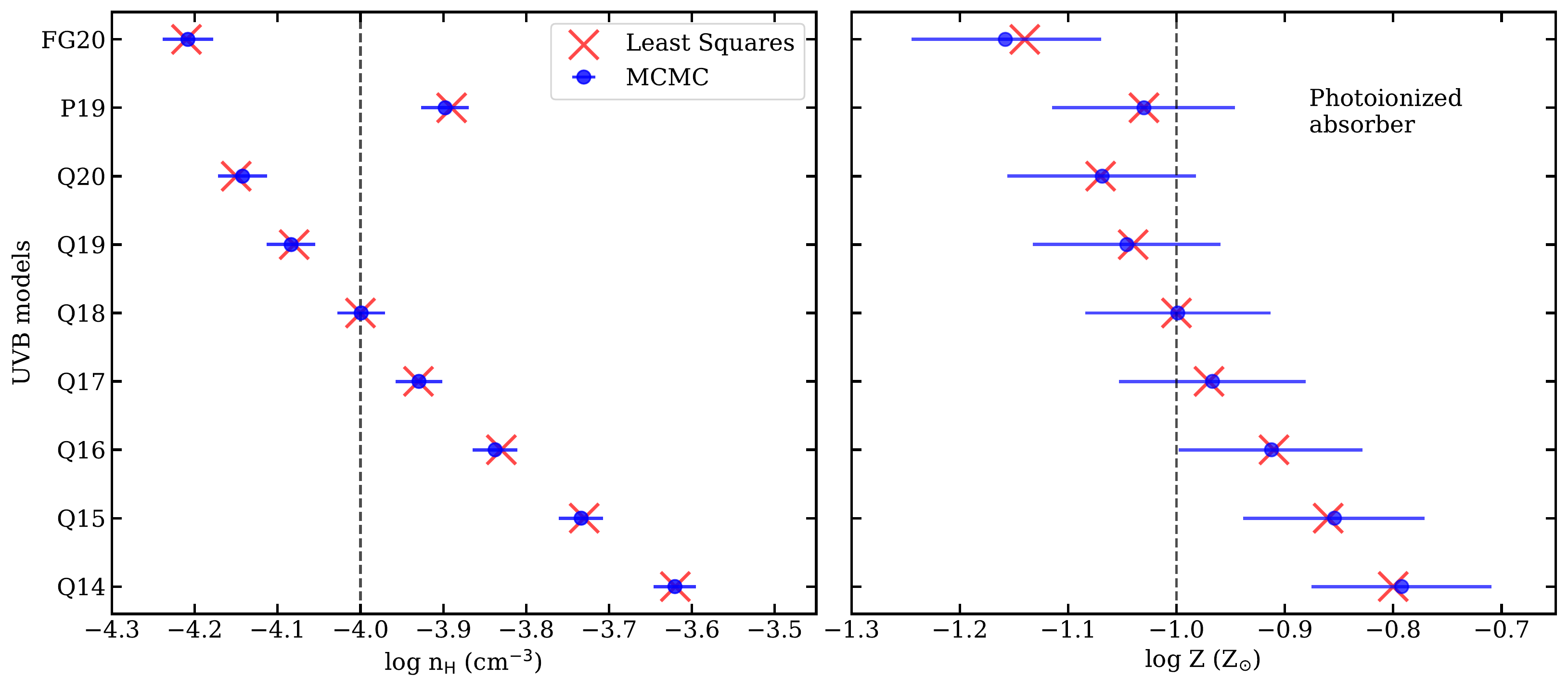}
\end{tabular}
\caption{ Inferred values of $n_{\rm H}$ (cm$^{-3}$; left-hand panel) and $Z/Z_{\odot}$; right-hand panel) with different UV background models 
(labeled on $y-$axis; see Fig.~\ref{fig.uvb}) for the test-case of a photoionized 
absorber. We use column densities of five ions
\{$N_{\rm C IV}$, $N_{\rm N V}$, $N_{\rm O VI}$, 
$N_{\rm Si IV}$, $N_{\rm Ne VIII}$\} for the inference. 
Models Q14-Q20 are from \citetalias{KS19}. 
The circles with horizontal error-bars show the result from the Bayesian 
MCMC method (median with 16 and 84 percentile values of posterior PDF;
see Fig.~\ref{figA} for few illustrations) while the red crosses shows the result 
of least square minimization method. Vertical dashed lines indicate the true values 
from toy observations which used Q18 
\citetalias{KS19} UV background model. 
The total difference in the inferred density  $\Delta_{\rm max} 
(\log n_{\rm H}) = 0.66$ and metallicity $\Delta_{\rm max} (\log Z) =0.48$ 
across all UV background models.
}
\label{fig.result_toy_photoionized}
\end{figure*}


First, we ran a huge grid of {\sc cloudy} simulations with an 
equally spaced dense grid of $\log n_{\rm H}$ (from $-5$ to $-2$ with 
$\Delta \log {n_{\rm H}} = 0.02$) and $\log Z$ (from $-3$ to $1$ with 
$\Delta \log Z = 0.05$). 
We repeat such a simulation grid for all nine UV background models.
For each UV background used in the {\sc cloudy} model, 
we stored the model predicted column densities of  metal ions 
\{$\log N^{\rm uvb}_{\rm ion}$\} for every input parameter 
$\theta^{\rm uvb} \equiv (\log n_{\rm H}, \log Z)$. 
Then, using methods described below we compare 
$\{\log N^{\rm uvb}_{\rm ion}\}$ to the 
$\{\log N^{\rm obs}_{\rm ion}\}$ in order to infer the parameters 
$\theta^{\rm  uvb}$. 

We developed two methods that use our toy observations (a set of
ion column densities $\{\log N^{\rm obs}_{\rm ion}\}$)
and our {\sc cloudy} model grid to infer the 
$n_{\rm H}$ and $Z$ for different assumed UV background models. 
First method uses
Bayesian formalism along with MCMC to explore the posterior probability density 
function (PDF) of parameters \nh~and \z. 
This method was used to mimic the real
observations where observed column densities have uncertainties.
To mimic that, we use errors
drawn from a Gaussian distribution on the mock observed column densities. 
In the second 
method, we determine $n_{\rm H}$  and $Z$ by simply minimizing the least square 
difference between the observed and modelled column densities of metal ions. 
We later show that, results of
both methods are in excellent agreement. These methods are described below.

In our first method, using Bayes theorem we write down the posterior PDF as  
\begin{equation}
\begin{split}
P(\{\log N^{\rm uvb}_{\rm ion}\} (\theta^{\rm uvb}) |  \{\log N^{\rm obs}_{\rm ion}\}) \propto  \\
P( \{ \log N^{\rm obs}_{\rm ion}\} | \{ \log N^{\rm uvb}_{\rm ion}\} (\theta^{\rm uvb}))P(\{ \log N^{\rm uvb}_{\rm ion}\} (\theta^{\rm uvb})),
\end{split}
\end{equation}
where we chose priors $P(\{ \log N^{\rm uvb}_{\rm ion}\} (\theta^{\rm uvb}))$ to be 
flat for both parameters $\log n_{\rm H}$ and $\log Z$. We use likelihood 
$\mathcal{L} \equiv P( \{ \log N^{\rm obs}_{\rm ion}\} | \{ \log N^{\rm uvb}_{\rm ion}\} (\theta^{\rm uvb})) $ to be Gaussian as

\begin{equation}
\mathcal{L} =  \prod_{i} \frac{1}{\sqrt{2 \pi} \sigma_i} 
\exp \Big( - \frac{(\log N^{\rm uvb}_{i}(\theta^{\rm uvb}) - \log N^{\rm obs}_{i})^2} {2 \sigma^{2}_i}  \Big).
\end{equation}
Here, index $i$ represents a metal ion from the set of ions chosen for the inference
(i.e the set of ions in mock observations)
and $\sigma _{i}$ is the error on $\log N^{\rm obs}_{i}$. For our toy observations we
generate random Gaussian errors $\sigma_{i}$ 
within a range $0.01 -0.25$ dex.\footnote{ We note that the 
inferred median of posterior does not depend on the values of 
the errors but the confidence interval
scales with them. Also, we just assign Gaussian errors to the column densities without altering their values.} 
We then use MCMC \citep[the emcee code by][]{emcee_paper} to sample the 
posterior PDF by interpolating between our dense {\sc cloudy}  model grids.

First we perform the inference test where we use the same Q18 \citetalias{KS19} 
UV background in {\sc cloudy} models 
to check how close our inferred values get to the true values. 
The results of such inference test are shown in the Fig. \ref{fig.inference}
for both photoionized absorber as well as $10^{5.5}K$ warm-hot absorber.
For both inferences, we use the same set of five metal ion column densities in observations; 
\{$N_{\rm C IV}$, $N_{\rm N V}$, $N_{\rm O VI}$, 
$N_{\rm Si IV}$, $N_{\rm Ne VIII}$\}.
 Fig. \ref{fig.inference} shows that 
the median of marginalized $\log n_{\rm H}$ and $\log Z$ values 
are in excellent agreement with true values $-4$ and $-1$, respectively. 
 
We reproduce the same results with a random choice of
any two or more metal ions from a large set of metal ions. We find that to jointly 
constrain the $n_{\rm H}$ and $Z$ we need to use at least two metal ions and they 
need not to be from the same element.

We find that when we use same UV background model as in the mock observations
the inferred median values of $n_{\rm H}$ and $Z$ always 
match with the true values (within $0.01$ dex) for both 
photoionized absorber and warm-hot absorber 
with $T \lesssim 10^{6.5} K$. For high temperature gas 
$T \gtrsim 10^{5.75}$ K, however, 
we need to reduce the error on our mock column densities to unrealistically 
small values to pass an inference test. 
Therefore, we restrict our-self to to the warm-hot
absorber models with $T  = 10^{5.5}$ K. 
However it is interesting to see how the joint posteriors of $n_{\rm H}$ and $Z$ changes 
with the $T$, which is explored in Fig.~\ref{figB} of the appendix.

Although the Bayesian MCMC approach is best suited for real observations which has 
Voigt profile fitting errors on column densities, the median values of parameters 
($n_{\rm H}$ and $Z$) can be simply obtained by minimizing the least 
square difference, denoted by LS here,
\begin{equation}
    LS(\theta^{\rm uvb}) = \sum_i (\log N^{\rm uvb}_{i}(\theta^{\rm uvb}) - \log N^{\rm obs}_{i})^2,
\end{equation}
between the observed and model column densities. The least square fitting approach is
also computationally least expansive. 
We confirm that we reproduce the true values of $n_{\rm H}$
and $Z$ using this method. 
Later, for quantifying the uncertainty  on the inferred 
 values of $n_{\rm H}$ and $Z$ from different assumed UV background models
 we use both methods, results of which are discussed in the next section.

\begin{figure*}
\begin{tabular}{cc}
  \includegraphics[width=0.48\textwidth]{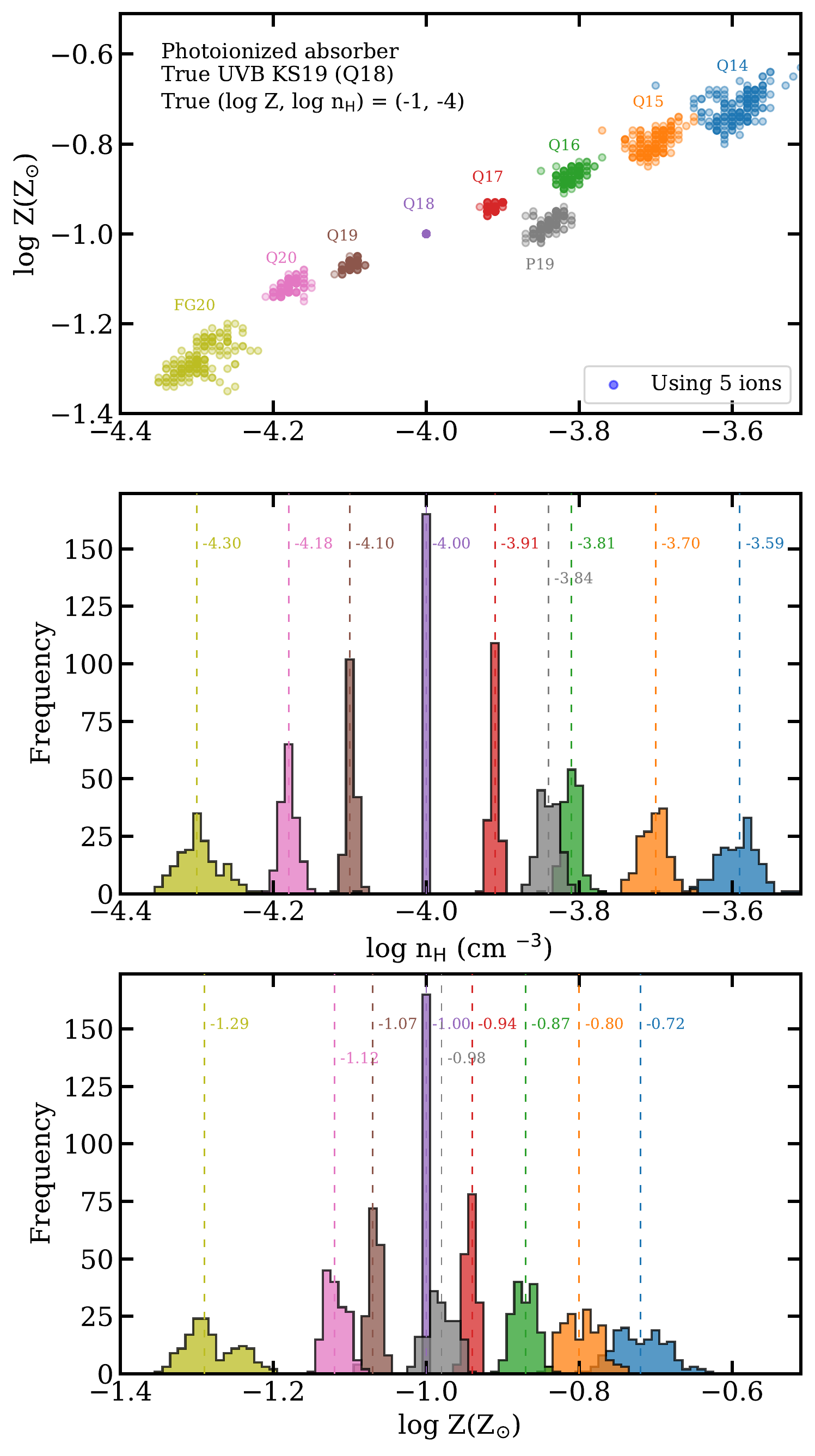} &
    \includegraphics[width=0.48\textwidth]{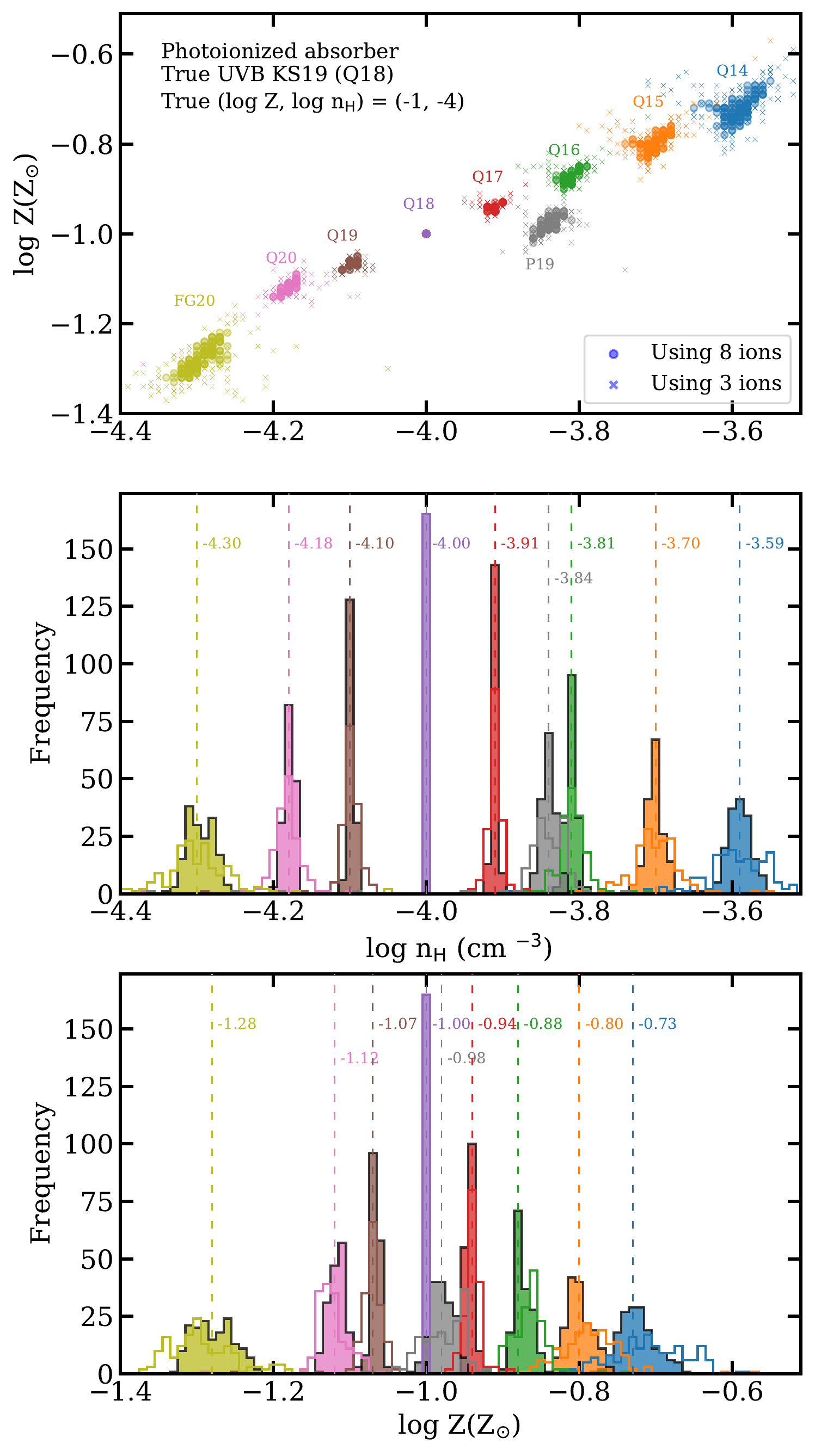}
\end{tabular}

\caption{ 
Distribution of inferred \nh~and \z~for photoionized mock absorber 
obtained using 165 sets of five ions
(left-hand panel), three ions and eight ions (right-hand panels.) 
Top panels show joint \nh~and \z~values whereas middle and bottom panels show
the histograms of \nh~and \z, respectively.  Different colors are used for different UV backgrounds. Vertical dashed lines and annotated values close to it show the median of the distributions. 
In both panels, true values of \nh~and \z~are 10$^{-4}$ cm$^{-3}$ and $Z = 10^{-1}$
$Z_{\odot}$ and Q18 \citetalias{KS19} UV background. 
In the right-hand side top panel shows \nh~and \z~inferred from three ions with crosses
and their distributions with open histograms at the middle and bottom panels, and from eight ions with circles and their distribution with solid histograms at the middle and bottom panels.
From this, it is evident that higher the number of ions, the better are the individual constraints of \nh~and \z.
Note that the width of individual histogram is significantly smaller that
the variation in \nh~and \z~arising from all UV background models. Here, 
\dnmax$=0.71$ dex and \dzmax$={0.56}$ dex (for all nine UV background models) 
when we use eight ions for the inference. 
}
\label{fig.photo_all_1}
\end{figure*}
\section{Results and Discussions}\label{sec:res}
In this section, we present the result of our analysis for both photoionized 
and warm-hot CGM gas. We first show the results for a test case 
where we use the same set of ions used for inference test 
(as shown in Fig.~\ref{fig.inference}) 
and then discuss results where we use different sets of ions
to quantify the distribution of inferred $n_{\rm H}$ and $Z$. 
We also discuss
the effect of shape and normalization of UV background spectra 
on the inferred \nh~and \z.

\subsection{Photoionized Absorbers}\label{sec:4.1}

\begin{figure*}
\begin{tabular}{cc}
    \includegraphics[width=0.48\textwidth]{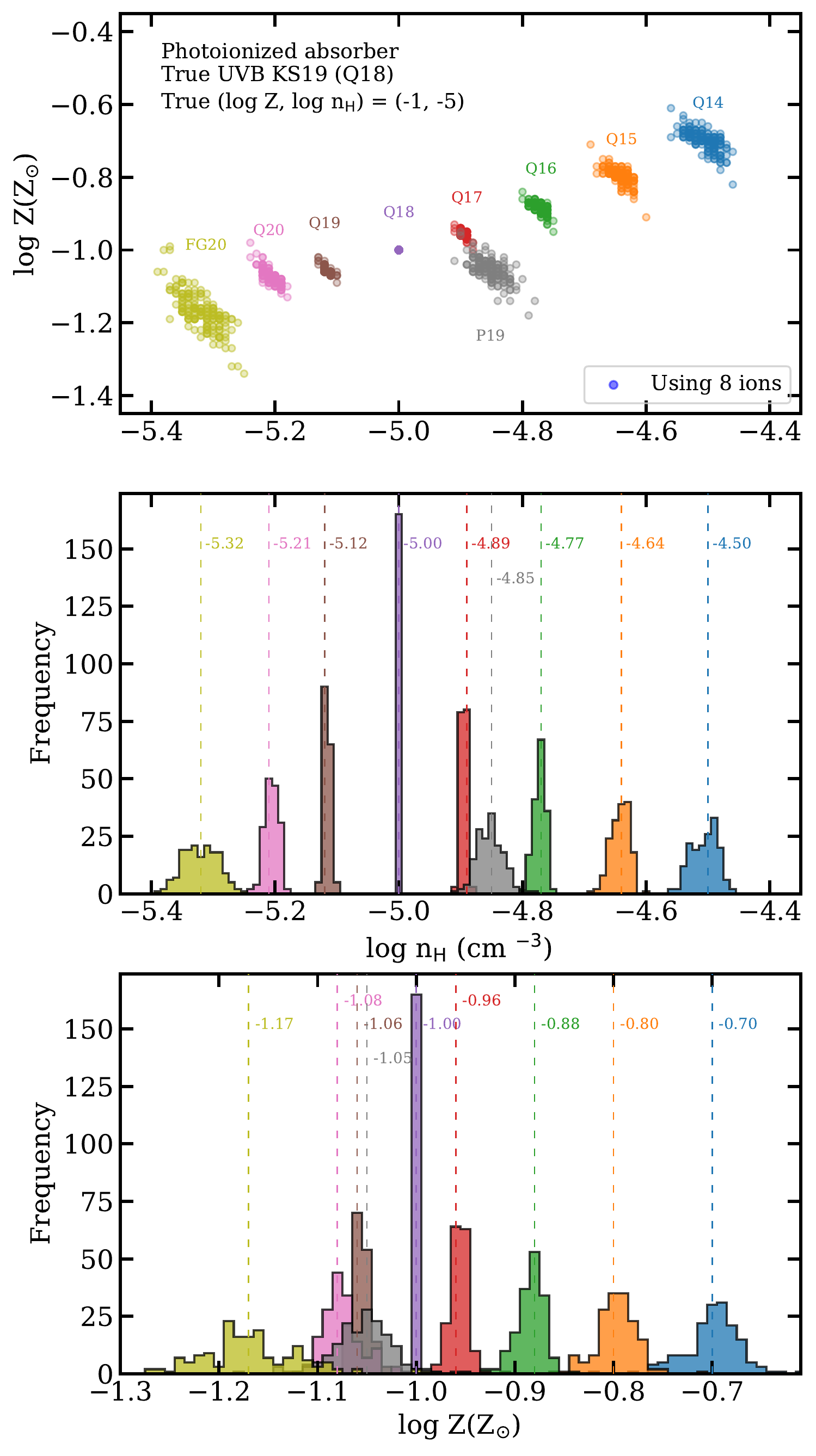} &
  \includegraphics[width = 0.48 \textwidth]{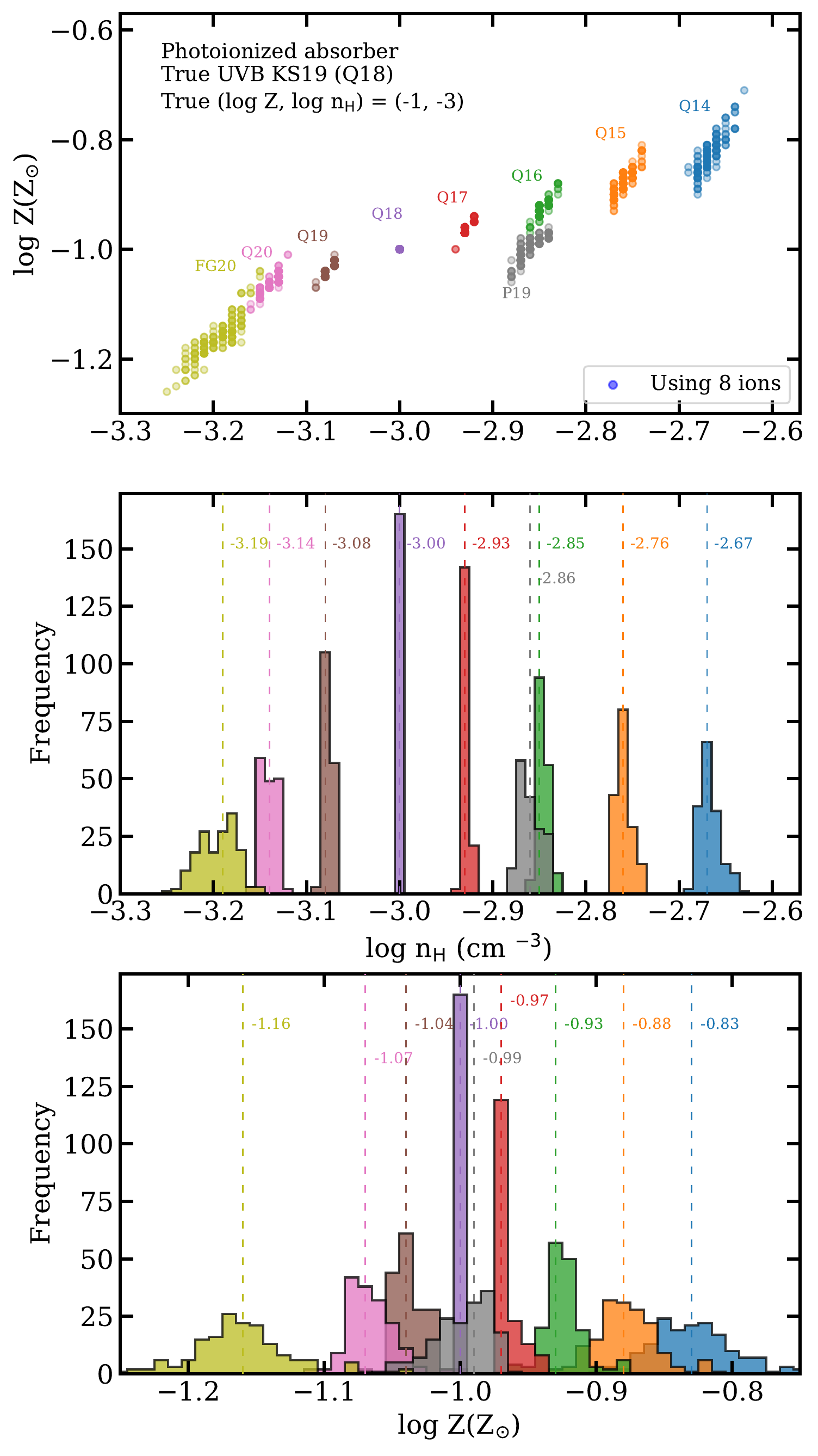}
\end{tabular}

\caption{
Same as Fig.~\ref{fig.photo_all_1} but for true value of $Z = 10^{-1} Z_{\odot}$
and \nh$=10^{-5}$ cm$^{-3}$ (left-hand panel) and  10$^{-3}$ cm$^{-3}$ (right-hand
panel) obtained with 165 sets of eight ions. Here, total variation \dnmax$= 0.82$ and 
\dzmax$=0.47$ in left-hand panel and \dnmax$= 0.52$ and 
\dzmax$=0.33$ in right-hand panel which indicates that with increasing density
(i.e higher true \nh) the variation
in the inferred \nh~and \z~arising from uncertain UV background decreases.}
\label{fig.photo_all_2}
\end{figure*}

\begin{figure}
\includegraphics[width=0.48\textwidth,height=\textheight,keepaspectratio]{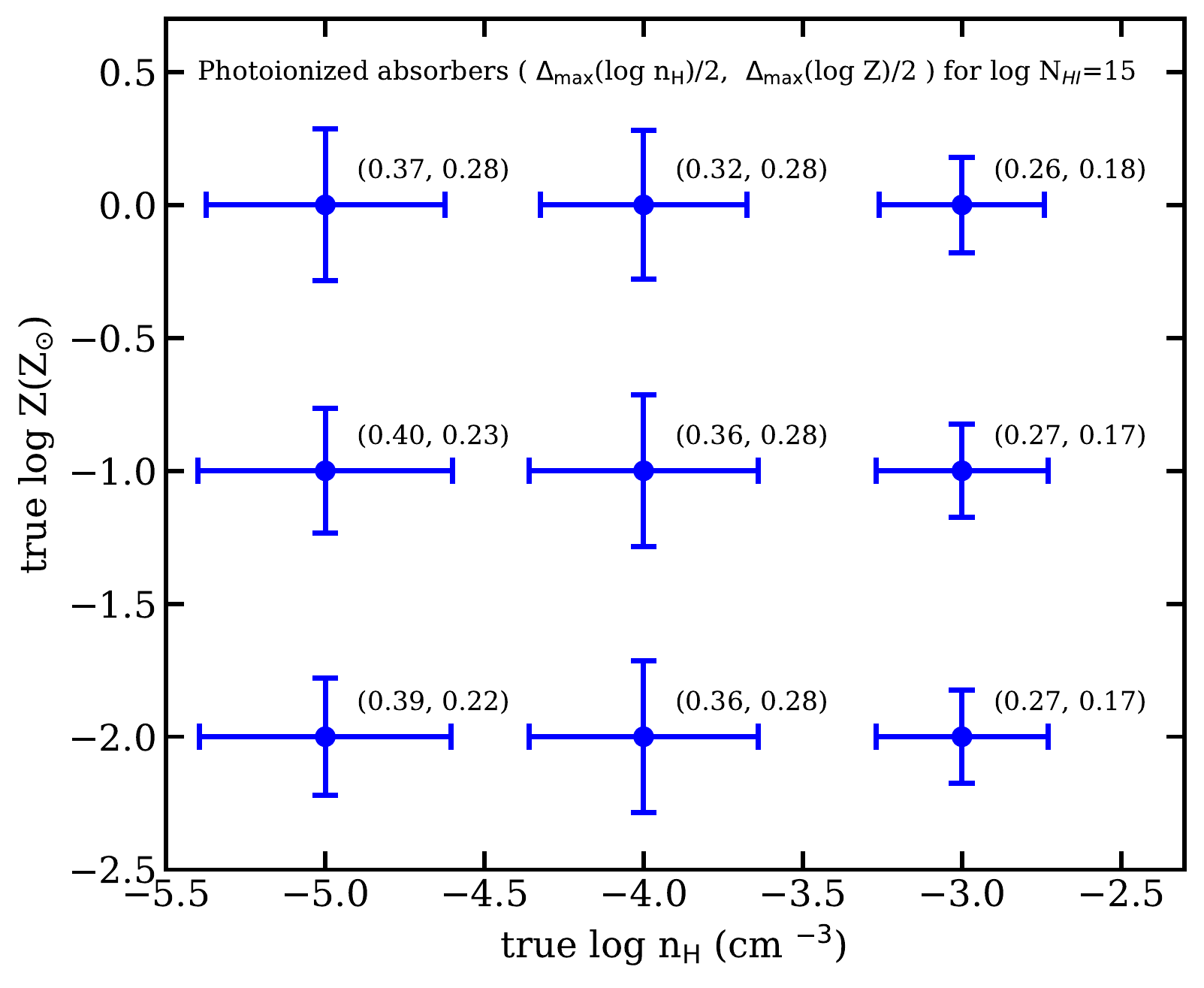}
\caption{
The grid of true \nh~and \z~where the extent of error-bars show the 
uncertainty (\dmax~values) on the  inferred \nh~(horizontal error-bars) and  
\z~(vertical error-bars) arising from uncertain 
UV background (using all nine UV background models) for photoionized gas. 
The annotated values in the brackets show the \dnmax/2 and \dzmax/2 which can
be quoted as systematic uncertainty on the inferred values of \nh~and \z~from
CGM observations for absorber with $N_{\rm HI}= 10^{15}$ cm$^{-2}$. 
These \dmax~values are also given in Table~\ref{tab1}.
See Fig.~\ref{fig.FigB} for the results obtained  for absorbers with $N_{\rm HI}$
ranging from from $10^{14}$ to $10^{19}$ cm$^{-2}$. }
\label{fig.res_photo_final}
\end{figure}

For a test case, inferred values of 
the $n_{\rm H}$ and $Z$ obtained using both Bayesian and least square 
methods by assuming different UV background  models are
shown in Fig.~\ref{fig.result_toy_photoionized} for our photoionized mock absorber.
For this particular example, we used the same set of five ions as we
used in the inference test (Fig.~\ref{fig.inference}). 
Median values of $n_{\rm H}$ and $Z$ from MCMC Bayesian fits match very well 
with the values obtained from least square minimization. 
We note that the size of confidence intervals on $n_{\rm H}$ and $Z$, 
i.e the 16 and 84 percentile values of the posterior PDFs, scales withr
the error on the metal-ion column densities, whereas the median values
of inferred $n_{\rm H}$ and $Z$ remain same. 
In Fig.~\ref{figA} of the appendix, we show a few example posterior 
PDFs for the Bayesian MCMC inference whose  results (marginalized $n_{\rm H}$
and $Z$ values with confidence intervals)
are shown in Fig.~\ref{fig.result_toy_photoionized}.
The expected variation in the inferred $n_{\rm H}$ and $Z$ values can be seen 
in Fig.~\ref{fig.result_toy_photoionized}. For photoionized models
the true $n_{\rm H}$ and $Z$ (shown by vertical dashed lines in 
Fig.~\ref{fig.result_toy_photoionized}) 
are reproduced only when the true UV background model,
Q18 \citetalias{KS19}, is used in the {\sc cloudy}.
The total variation in the inferred $n_{\rm H}$ is 0.57 dex across
the seven UV background models of \citetalias{KS19} and 0.66 dex if
we include remaining two UV background models. 
Whereas variation in $Z$ is less as compared to variation in $n_{\rm H}$,
it is 0.35 dex for seven \citetalias{KS19} models and 0.48 dex 
across all nine UV background models. The inferred values of
$n_{\rm H}$ and $Z$ for \citetalias{Puchwein19} lie close to the 
inferred values for Q16-Q18 models of \citetalias{KS19}. A possible reason 
for this is discussed in section \ref{sec.shape_and_norm}. 
The softest UV background \citetalias{FG20} provides lowest values of 
\nh~and \z. Fig.~\ref{fig.result_toy_photoionized} shows a clear trend that 
\emph{the softer UV background models 
(refer to Fig.~\ref{fig.uvb}) leads to the lower
values of inferred $n_{\rm H}$ and $Z$}. It is especially apparent in
seven model of \citetalias{KS19} where the shape of UV background depends 
only one one parameter i.e the intrinsic SED ($\alpha$) of QSOs.
Such a trend can be naively explained in terms of the ionization 
parameter $U$. Because for optically thin cloud 
the abundance pattern of ions depends on the ionization parameter 
$U = n_{\gamma}/ n_{\rm H}$ of hydrogen, where number density of hydrogen ionizing 
photons $n_{\gamma}$ reduces with increasing $\alpha$.
Therefore, for softer UV background with higher $\alpha$ the inferred $n_{\rm H}$ 
decreases to preserve $U$ in order to 
reproduce ion column densities as in mock observations.

To understand the dependence of our inferred \nh~and \z~on the 
choice of different ions used in the mock observations, we use a large number of 
different sets of ions and infer \nh~and \z. For that, we first identify ions whose 
column densities are more than $10^{11}$ cm$^{-2}$ in a {\sc cloudy} model ran with 
$n_{\rm H} = 10^{-4}$ cm$^{-3}$ and $Z = 0.1$ $Z_{\odot}$ with true UVB as 
\citepalias[][]{KS19} Q18 out of 23 different metal ions commonly observed in the 
CGM across all redshifts.\footnote{The 23 ions considered are: C II, C III, C IV, N III, N IV, N V, O II, O III, O IV, O V, O VI, O VII, S II, S III, S IV, S V, S VI, Si II, Si III, Si IV, Mg II, Ne VIII and Fe II.}
We select minimum column density  threshold $10^{11}$ cm$^{-2}$ to avoid very low 
columns of ions produced in {\sc cloudy} models and verified that our final results 
are not sensitive to this threshold as long as it is larger than  $10^{11}$ cm$^{-2}$. Then, out of these selected ions we create
all possible sets of different ions with varying number of ions in each set. We then
group these sets according to number of ions they have, from two ions to eight ions.
From such sets with a fixed number of ions we randomly pick 165 unique 
sets.\footnote{The number 165 is arrived at by firstly choosing the 
minimum number of ions that satisfy the above screening criteria, with the stopping 
criteria set to $N_{\rm HI} = 10^{14.5}$ cm$^{-2}$ which is found to be 11. Then, 
the number of largest possible sets of 8 ions each is $^{11}$C$_8$ = 165.} Next, we
use column densities of these ions (from our toy CGM absorber) in each set and treat
it as one mock observation to infer the \nh~and \z~assuming all nine UV background 
models. For such inferences we use the least square minimization method since it is 
computationally efficient and the results agree very well with the Bayesian MCMC 
method.

The distributions of the inferred \nh~and \z~are shown in 
Fig.~\ref{fig.photo_all_1} for all nine UV background models
where our mock CGM absorber had  
true values of $n_{\rm H}  =10^{-4}$ cm$^{-3}$ and 
$\log Z = -1$
with Q18 \citepalias{KS19} as a true UV background.
Top panels of Fig.~\ref{fig.photo_all_1} show inferred values of 
\nh~and \z~for 165 sets of five 
(left-hand panel), three and eight (right-hand panel) metal ions each. 
Middle and bottom panels show the distributions of inferred \nh~and \z~for
each UV background model. First thing to note that no matter how many ions
and which set of ions we use, we always get back the true \nh~and \z~when 
we use true UV background model in {\sc cloudy}, 
as evident from $\delta$-function like
histograms for Q18 models. When we use different UV background model,
we see that different set of ions give slightly different results. The scatter
in these results widens the distribution of inferred 
\nh~and \z, especially apparent
when we use the UV background that is quite different than the true 
UV background. It is rather interesting to note that distributions of
\nh~and \z~widens monotonically as we move to softer or harder UV 
background models, except for \citetalias{Puchwein19}. 
This interesting trend is apparent in the results obtained for all sets of ions
with different numbers, be it three ions or eight ions (see right-hand panel of
Fig.~\ref{fig.photo_all_1}). Such a trend can be
used to constrain the true model of UV background 
by using CGM observations  
which we plan to explore in our future work. 

Fig.~\ref{fig.photo_all_1} also shows the expected trend that when we use 
more ions, e.g eight as compared to three as shown in right-hand panel of the
figure, the constraints on \nh~and \z~become better. In other words, 
the scatter around the median inferred value of \nh~and \z~reduces 
when one uses sets of ions with a larger number of ions for the inference. 
This can be seen from right-hand panels of 
Fig~\ref{fig.photo_all_1} where filled histograms obtained for 165 sets of eight ions are 
narrower than the the open histograms obtained for 165 sets of three ions.
However, note that this scatter around the median value (i.e the width of 
distribution) for both inferred \nh~and \z~is significantly 
lower as compared to the differences in
the median values for the range of UV background models, 
even when we use sets of just three ions. Therefore the uncertain UV background
can have major contribution in the total uncertainty of the inferred \nh~and \z.
Dotted lines in the middle and bottom panels indicate the median values of
the inferred \nh~and \z~which are also annotated on the figure with same
colors. We note that the median values converge within 0.01 dex when we use more 
than four ions to infer mean \nh~and \z, but even if we use only 
three or four ions the median values differ by less than 0.05 dex 
from the converged values.
We do not recommend using only two ions, unless they are from the same
element, because in that case the scatter in the inferred values of \nh~and 
\z~can be of the order of a magnitude. 

In Fig.~\ref{fig.photo_all_1}, we see the same trends in the median values 
of inferred \nh~and \z~as in the test case shown in 
Fig~\ref{fig.result_toy_photoionized}, i.e the softer UV background yields
lower \nh~and \z.
To quantify the total variation in the inferred $n_{\rm H}$ and $Z$ values
across different UV background models
we define $\Delta_{\rm max} (\log n_{\rm H})$ and $\Delta_{\rm max} (\log Z)$.
These quantities represent the expected variation
in the $n_{\rm H}$ and $Z$ due to uncertain UV background.
The values of $\Delta_{\rm max} (\log n_{\rm H})$ and 
$\Delta_{\rm max} (\log Z)$ are obtained
by taking the difference in the median of 
inferred $n_{\rm H}$ and $Z$ values for each 
UV background model and then finding out the maximum of that difference.
Given the trend of inferring lower values of $n_{\rm H}$ and $Z$ with 
softer UV background, the $\Delta_{\rm max}$ values 
when quoted for all nine UV background models,
are just the 
differences between median inferred $n_{\rm H}$ and $Z$ obtained for 
Q14 \citetalias{KS19} and \citetalias{FG20} models. 
Whereas, when we quote the $\Delta_{\rm max}$ values for only seven 
\citetalias{KS19} UV background models, they represent the difference between
inferred $n_{\rm H}$ and $Z$ obtained for 
Q14 and Q20 models of \citetalias{KS19}. 

From Fig.~\ref{fig.photo_all_1}, 
we find that $\Delta_{\rm max} (\log n_{\rm H}) = 0.59$ dex and  
$\Delta_{\rm max} (\log Z) = 0.39$ dex 
for the seven UV background models of \citetalias{KS19}. 
For all nine UV background models, we find 
$\Delta_{\rm max} (\log n_{\rm H}) = 0.71$ dex and  
$\Delta_{\rm max} (\log Z) = 0.56$ dex. 
These results are obtained by using sets of eight ions for the inference
(see right-hand panel of Fig.~\ref{fig.photo_all_1}), the true values of 
\nh~and \z~as $10^{-4}$ cm$^{-3}$ and $0.1 Z_{\odot}$, and  
for true Q18 \citetalias{KS19} UV background. In order to check the dependence of
different values of true \nh~and \z~used for creating the mock CGM absorber on the 
values of \dmax, we redo the analysis for nine combinations of true 
\nh~and \z~where we change \nh~from 10$^{-5}$ to 10$^{-3}$ cm$^{-3}$
and \z~from $0.01$ to 1 Z$_{\odot}$. Results for two such combinations (log \z~ = -1 and log \nh~ = -3, log \z~ = -1 and log \nh~ = -5)
are shown in Fig.~\ref{fig.photo_all_2} where the \nh~and \z~inference 
used 165 sets of eight ions. 
The \dmax~values differ for these two cases by a factor of $\sim 0.2$ dex.

\begin{figure*}
\begin{tabular}{cc}
  \includegraphics[width=0.82\textwidth]{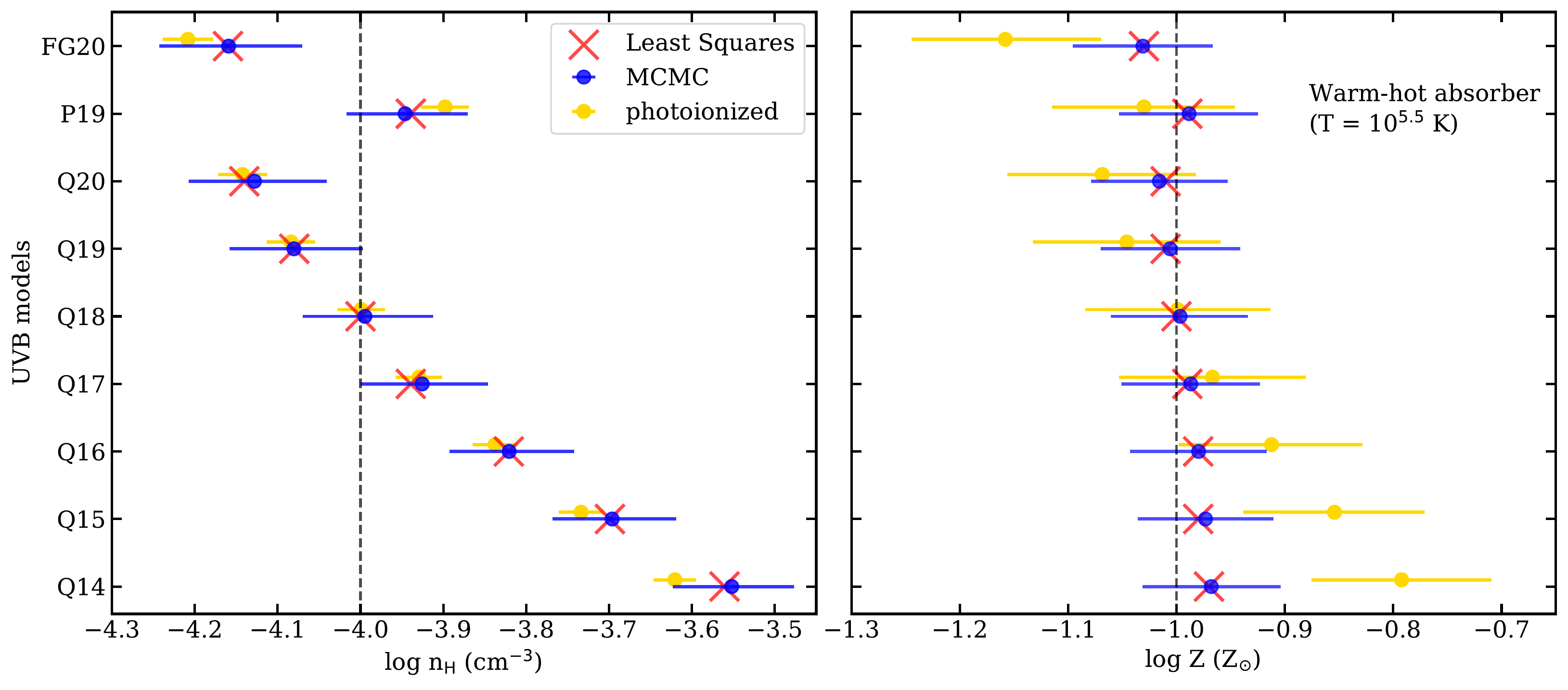} 
\end{tabular}

\caption{ Inferred values of $n_{\rm H}$ (cm$^{-3}$; left-hand panel) and $Z$ 
(Z$_{\odot}$; right-hand panel) 
with different UV background models (labeled on $y-$axis; see Fig.~\ref{fig.uvb}) 
for the the warm-hot absorber with $T= 10^{5.5}$ K. 
Models Q14-Q20 are from \citetalias{KS19}.
The circles with horizontal error-bars show the result from the Bayesian MCMC method 
(median with 16 and 84 percentile values of posterior PDF)
while the red crosses shows the results from least square minimization method. Vertical 
dashed lines indicate the true values from toy observations generated using Q18 
\citetalias{KS19} UV background model. The total difference in the inferred density 
$\Delta_{\rm max} (\log n_{\rm H}) = 0.55$ while the metallicity does not 
show any significant deviation ($\Delta_{\rm max} (\log Z) =0.03$) from the true value.
Yellow color data points show the result for photoionized absorber test case
(from Fig.~\ref{fig.result_toy_photoionized}) for comparison.
}
\label{fig.result_toy_hybrid}
\end{figure*}
We find that for the same true \nh~and \z~there is a small change in
\dmax~values of the order of 0.1 dex if we use different true UV 
background models, instead of fiducial Q18 \citetalias{KS19} model, 
in our mock CGM absorbers (as described in sec.~\ref{sec.toy_mode}). 
Therefore, we repeat the analysis
by changing the true UV background models and calculate \dmax~values.
We do this for all nine UV background models and take the mean of
the \dmax~values.\footnote{Note that we take mean for just 
removing the dependence on a true UV background, 
so that  one can quote $\pm$\dnmax/2 as an average systematic 
error from UV background irrespective of their choice of UV background used 
for inference (but see Section~\ref{sec:4.4}).} These mean values 
of \dnmax~and \dzmax~are given in Table~\ref{tab1} 
and also shown in Fig.~\ref{fig.res_photo_final} with
error-bars on the true values of \nh~and \z. The full extent of 
error-bars in the figure represent the values of \dmax. The annotated values 
in the figure are the \dnmax/2 and \dzmax/2 which is equivalent to 
being a systematic uncertainty on the log \nh~and log \z~arising from 
uncertain UV background. 
Fig.~\ref{fig.res_photo_final} shows that \dnmax~and \dzmax~has very 
weak dependence on true value of \z~but both reduce by $\sim 0.2$ dex when 
true \nh~changes from
10$^{-5}$ to 10$^{-3}$ cm$^{-3}$. The \nh~and \z~are better
constrained for gas at high densities than at low densities. 
This is mainly because at high densities self-shielding of gas 
reduces the impact of UV background.

Till now we have assumed that the absorber 
is optically thin with hydrogen column density $N_{\rm HI} =  10^{15}$ cm$^{-2}$ which is not true for most of the CGM absorbers. Therefore
we repeat the whole analysis with another sets of {\sc cloudy} models by varying $N_{\rm HI}$
from $10^{14}$ to $10^{19}$ cm$^{-2}$. The results of which, similar to the one shown in the Fig.~\ref{fig.res_photo_final} are summarized in the Fig.~\ref{fig.FigB} in Appendix. 

Overall, for the absorbers in photoionization equilibrium, 
the inferred \nh~can be uncertain up to a value of \dnmax~which ranges from
from 0.5 to 0.8 dex and 
inferred \z~can be uncertain up to a value of \dzmax~that ranges from 
0.2 to 0.6 dex for the gas with density 10$^{-3}$ to 10$^{-5}$ cm$^{-3}$ for
clouds with $N_{\rm HI}$
ranging from $10^{14}$ to $10^{19}$ cm$^{-2}$ 
(see Fig.~\ref{fig.FigB}).
Table~\ref{tab1} provides values of \dmax~for 
different true \nh~and \z~for all nine UV background (shown in 
Fig~\ref{fig.res_photo_final})  as well as for only seven \citetalias{KS19} UV 
background models for absorber with  $N_{\rm HI} = 10^{15}$  cm$^{-2}$.

\subsection{Warm-hot Absorber}\label{sec:4.2}
\begin{figure*}
\begin{tabular}{cc}
      \includegraphics[width=0.48\textwidth]{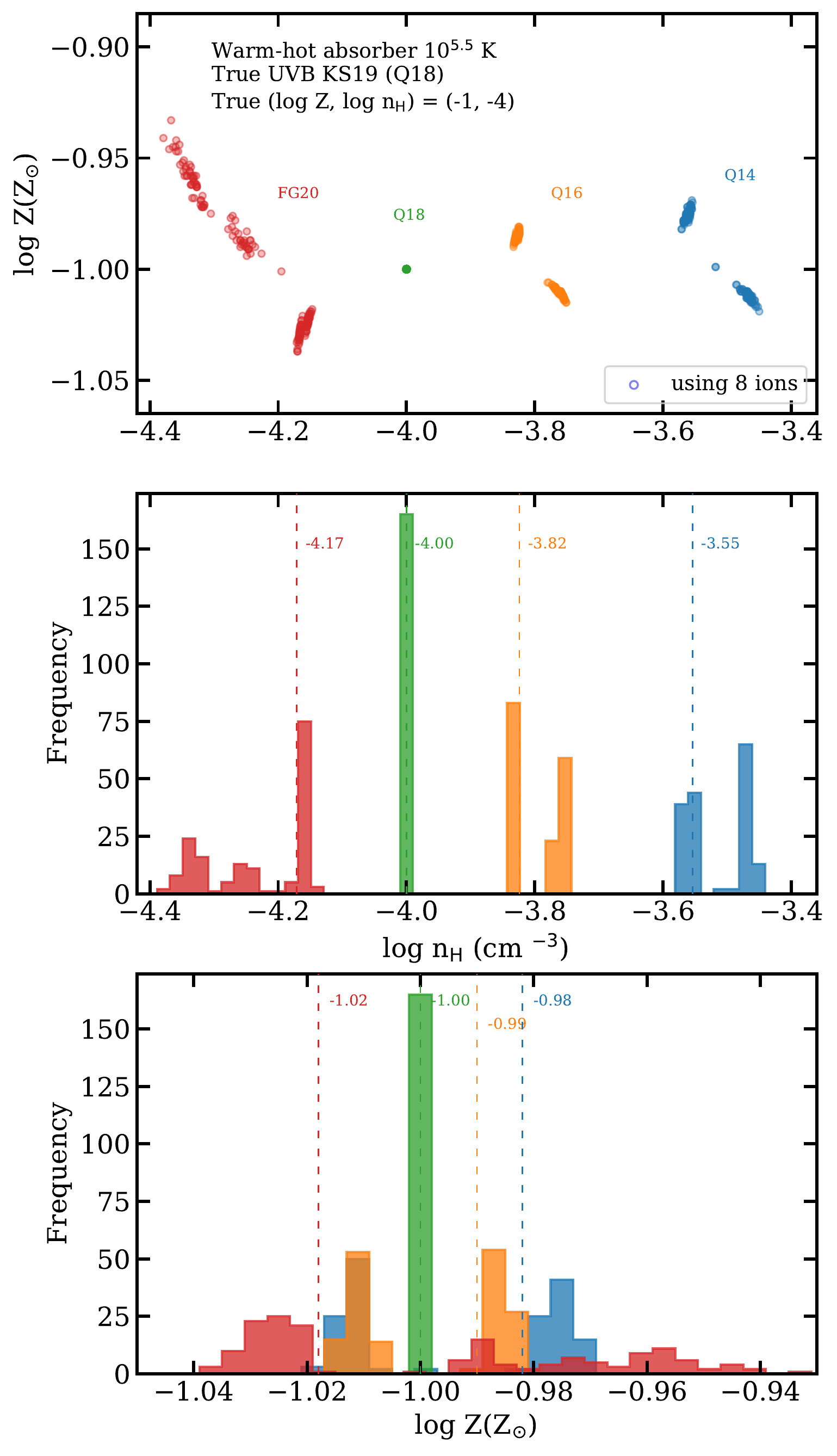} &
      \includegraphics[width = 0.48 \textwidth]{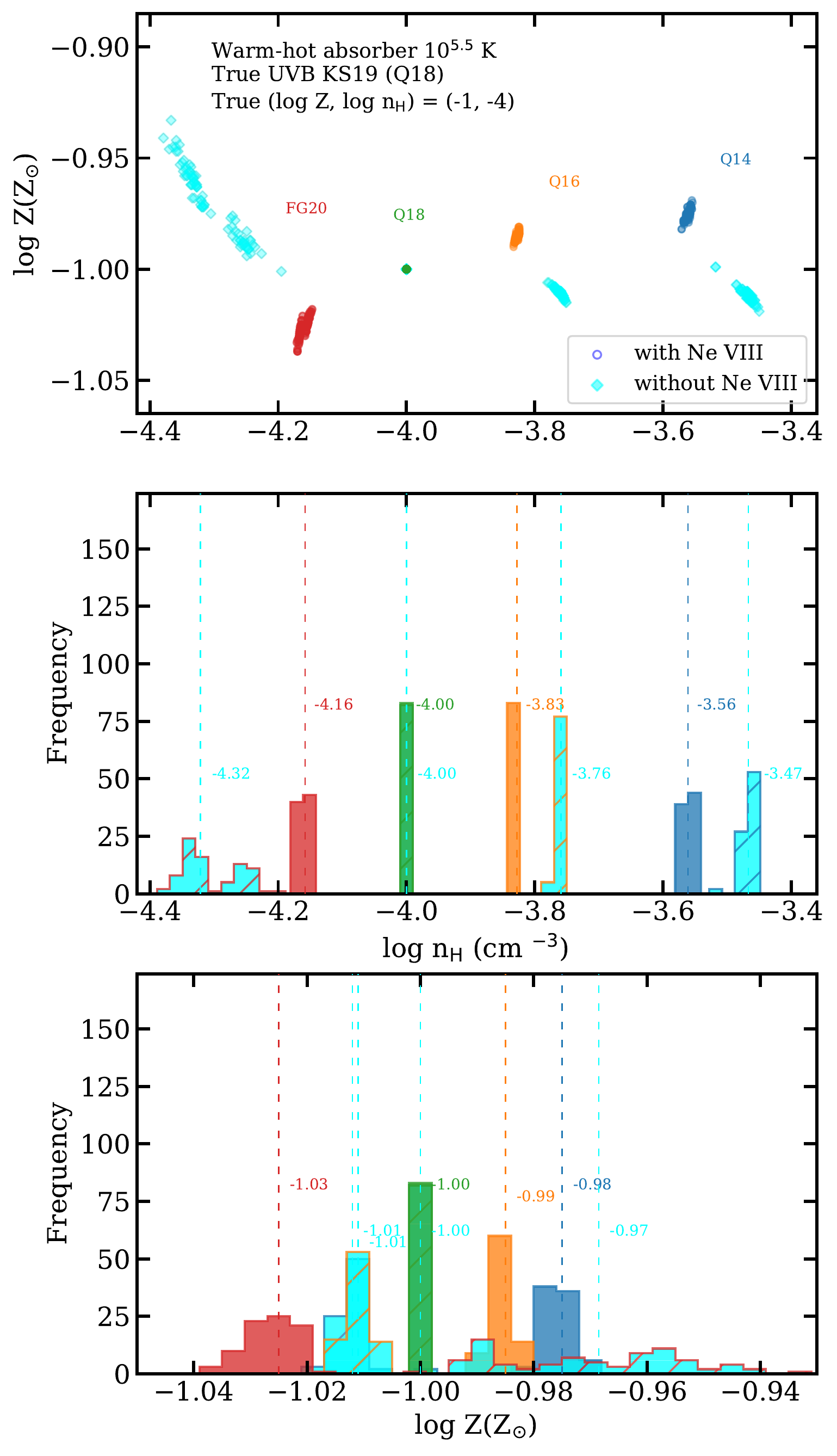}
\end{tabular}
\caption{ \emph{Left-hand panels}: 
Distribution of inferred \nh~and \z~for warm-hot mock absorber
at $T = 10 ^{5.5}$ K obtained using 165 sets of eight ions.
Top panels show joint \nh~and \z~values whereas middle and bottom panels show
the histograms of \nh~and \z, respectively.  Different colors are used for
different UV backgrounds. Vertical dashed lines and annotated values close to those are the median values of the distributions. 
There is a clear bi-modality in the inferred \nh~and \z~values, 
except when we use the true UV background model.
\emph{Right-hand panel}: Same as left-hand panel but cyan points
and histogram show the inferred \nh~and \z~obtained using set of 
ions not having Ne~{\sc viii} 
(82 out of 165). All the cyan points fall on one side of bi-modal 
distribution indicating inclusion or exclusion of Ne~{\sc viii} gives rise to the 
bi-modality. Moreover, inferred \nh~is closer to true value when Ne~{\sc viii} is 
included in the inference.  Total variation in \nh~is 0.6 dex when Ne~{\sc viii} 
is included in the inference and 0.85 dex otherwise. 
Note that, irrespective Ne~{\sc viii}, inferred \z~is close to true \z~within 0.03 dex. }
\label{fig.hybrid_all_1}
\end{figure*}
\begin{figure*}
\begin{tabular}{cc}
      \includegraphics[width=0.48\textwidth]{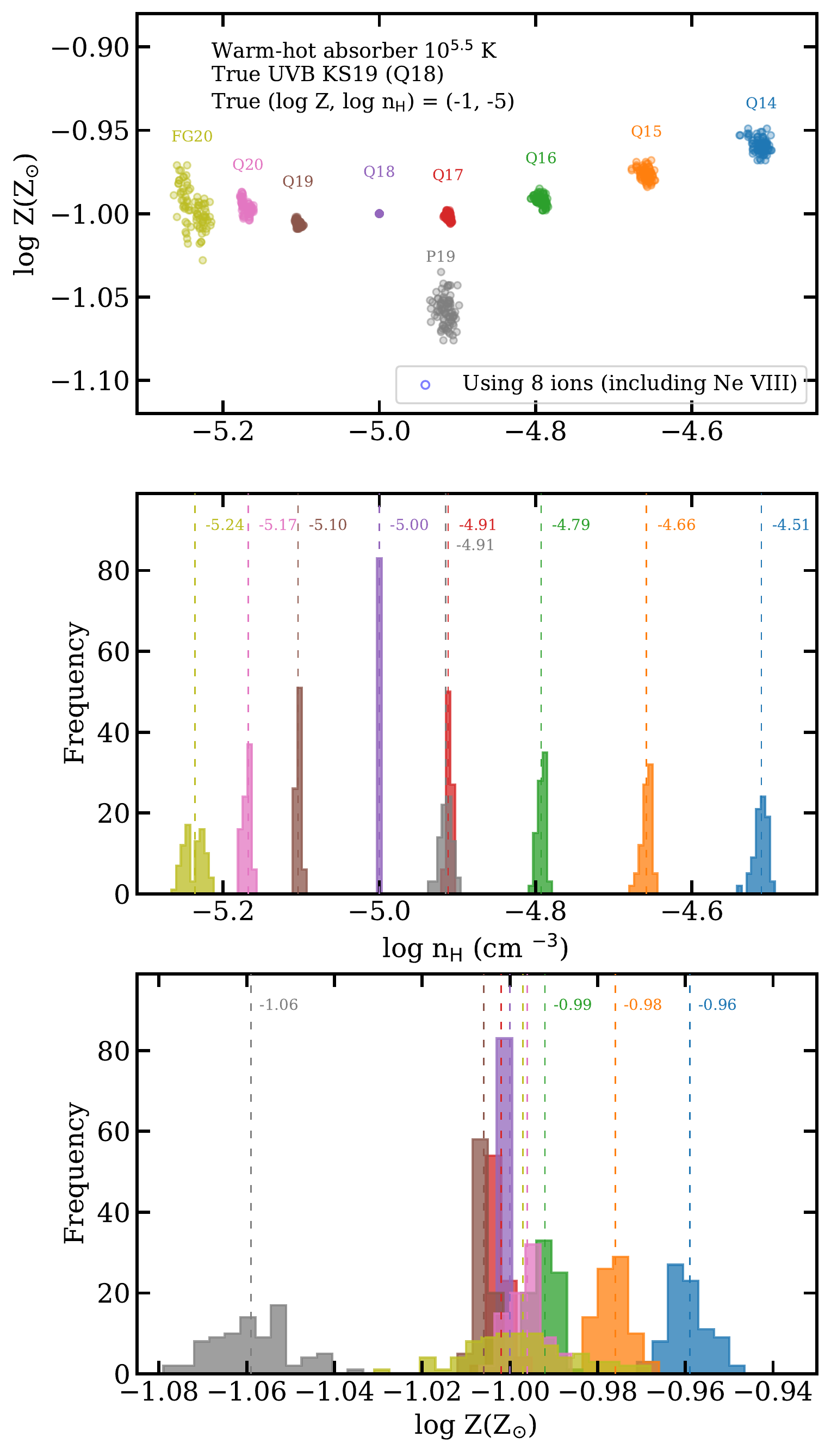} &
      \includegraphics[width =0.48\textwidth]{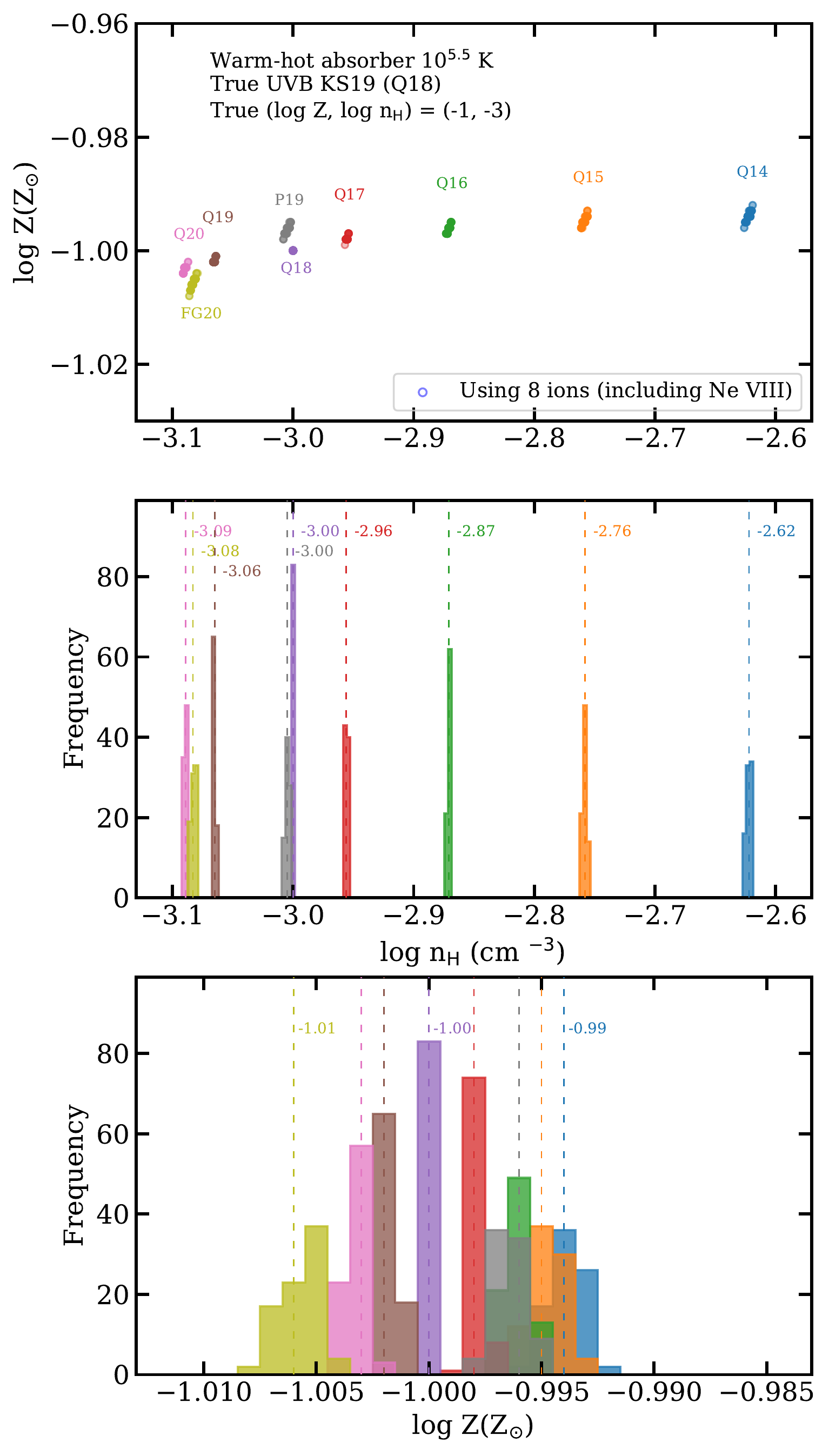}
\end{tabular}

\caption{ Same as the left-hand panel of Fig.\ref{fig.hybrid_all_1} but inferred \nh~and \z~shown 
only for 83 sets of ions out of 165 that have Ne~{\sc viii}. Left-hand panel show results
for true \nh$=10^{-5}$ cm$^{-3}$ and right-hand panel shows results for true 
\nh$=10^{-3}$ cm$^{-3}$. For both cases true \z$=10^{-1} Z_{\odot}$. We find 
\dnmax$=0.73$ and \dzmax$=0.1$ dex for left-hand panel and  \dnmax$=0.47$ and \dzmax$=0.02$ 
dex for right-hand panel indicating inferred \nh~and \z~show less variation at higher true \nh.
Given negligibly small deviation in inferred \z~from its true value, we can safely say that
estimates of \z~are robust for warm-hot absorbers.
 }
\label{fig.hybrid_all_2}
\end{figure*}
In Fig.~\ref{fig.result_toy_hybrid}, we show the 
inferred values of the $n_{\rm H}$ and $Z$ using both Bayesian MCMC and least square 
methods for a test case of a warm-hot absorber at constant temperature of 10$^{5.5}$ K. 
For the inference, we used the same set of five ions
used in the inference test (Fig.\ref{fig.inference}) and the test case of a 
photoionized absorber (Fig.~\ref{fig.result_toy_photoionized}). 
As in the photoionized case, median values of $n_{\rm H}$ and $Z$ from MCMC Bayesian 
fits match well with the values obtained from least square minimization.
For comparison, we also show the Bayesian MCMC results obtained for the 
photoionized absorber (as shown if Fig.~\ref{fig.result_toy_photoionized})
with yellow data points. We get almost same \nh~values as in case of the
photoionized absorber. It suggests that the extra ionization of gas  
contributed by UV background in addition to collisional ionization is responsible
for the change in inferred \nh. 
The total variation in \nh~is 0.56 dex, just 0.1 dex smaller than the
one obtained for test case of photoionized absorber.
However, constraints on \nh~become much weaker, as 
revealed by the extent of confidence intervals on blue and 
yellow points. We see similar trend in Bayesian MCMC posteriors at higher temperatures
where the constraints become weaker for \nh~and one needs to reduce the 
uncertainty on column densities of metals to get meaningful constraints. 
In Fig.~\ref{figB} of appendix, we show few example posteriors 
obtained for the inference test of warm-hot absorber 
up to $T = 10^{6.5}$K.

Inferred \z~values for this test case of  warm-hot absorber are
consistent with the true \z~and show very little variation across 
all UV background (maximum of 0.03 dex). This suggests that
the effect of collisions dominates over the effect of photoionization for metals. 
The constraints on \z~are slightly better than the one obtained for 
photoionized absorber, as can be seen from
the extent of confidence interval. 

To understand the dependence of our inferred \nh~and \z~on the choice 
of different ions,
we follow the same procedure as in case of photoionized absorber and
create 165 sets of ions  with a fix number of ions in each set. We use 
these sets for the inference where the column densities of ions in each set
obtained from our toy CGM warm-absorber serve as an individual mock observation. 
We use least square minimization for the inference. 
Results of which, obtained with sets of eight ions each, 
are shown in the left-hand panel of Fig.\ref{fig.hybrid_all_1}. 
For clarity we only show results obtained for
Q14, Q16, Q18 of \citetalias{KS19} and \citetalias{FG20} UV 
background models. Here, we have true \nh$= 10^{-4}$ cm$^{-3}$ 
and \z$= 10 ^{-1} Z_{\odot}$ with Q18 \citetalias{KS19} as a true UV 
background. Top panel shows individual \nh~and \z~values whereas middle and 
bottom panel shows
the histograms of those. When we use same Q18 \citetalias{KS19} UV background 
we reproduce true values all the time, as can be seen from the 
$\delta$-function like histograms for \nh~and \z.
For other UV background models, not all sets of ions infer the same 
\nh~and \z~resulting into wider distributions. However, unlike photoionized 
absorber we find that the distributions of inferred \nh~and \z~are bi-modal 
for UV background models apart from the true Q18 UV background. 
This is more clear for the Q14 and Q16 UV background in the 
left-hand middle panel of Fig.~\ref{fig.hybrid_all_1}. 
We found that this bi-modality is related to ion Ne~{\sc viii}. 
If we split all 165 sets of ions
in two groups, one group including Ne~{\sc viii} (total of 83 sets) 
and other without it (82 sets), inferred
values of \nh~from these two groups clearly fall on two sides of the histogram 
explaining the bi-modal nature of the original distribution. 
This is illustrated in right-hand panel of Fig.~\ref{fig.hybrid_all_1}
where \nh~and \z~inferred excluding Ne~{\sc viii} ion are shown 
with cyan color. Note that there is no bi-modality in inferred \nh~and 
\z~when we use the same true UV background model. 
Most interestingly, the set of ions having Ne~{\sc viii} column densities 
always infer the \nh~values close to true values as can be seen from 
the right-hand  middle
panel of the Fig.~\ref{fig.hybrid_all_1}. Without Ne~{\sc viii}, the true \nh~can
be off by factor of 0.1-0.16 dex. 
From Fig.~\ref{fig.hybrid_all_1}, 
we find that $\Delta_{\rm max} (\log n_{\rm H}) = 0.85$ dex 
without including Ne~{\sc viii} and  0.6 dex when including  Ne~{\sc viii}
for all nine UV background models. 
Inferred metallicity, being very close to
true values, does not show any significant dependence on the Ne~{\sc viii}. 
In both cases, with and without Ne~{\sc viii}, \dzmax~is very small 
($<0.05$ dex).
Note that for our test case shown in Fig.~\ref{fig.result_toy_hybrid} the 
set of five ions used for the inference has Ne~{\sc viii}.

In order to check the dependence of
different values of true \nh~and \z~used in mock warm-hot absorber on the the 
values of \dmax, as in the case of photoionized absorber
we redo the analysis for nine combinations of true 
\nh~and \z~where we change \nh~from 10$^{-5}$ to 10$^{-3}$ cm$^{-3}$
and \z~from 10$^{-2}$ Z$_{\odot}$ to 1 Z$_{\odot}$. Results for two such combinations
are shown in Fig.~\ref{fig.hybrid_all_2} where the \nh~and \z~inference 
used 83 sets of eight ions which include ion Ne~{\sc viii}. 
There is more scatter at the inferred values of \nh~and \z~for lower values of
true \nh, as can be seen by comparing left and right-hand panel of 
Fig.~\ref{fig.hybrid_all_2}. 

However, note that warm-hot absorbers show significantly less variation in the inferred $Z$ 
as compared to the photoionized absorbers.

\begin{figure}
\includegraphics[width=0.48\textwidth,height=\textheight,keepaspectratio]{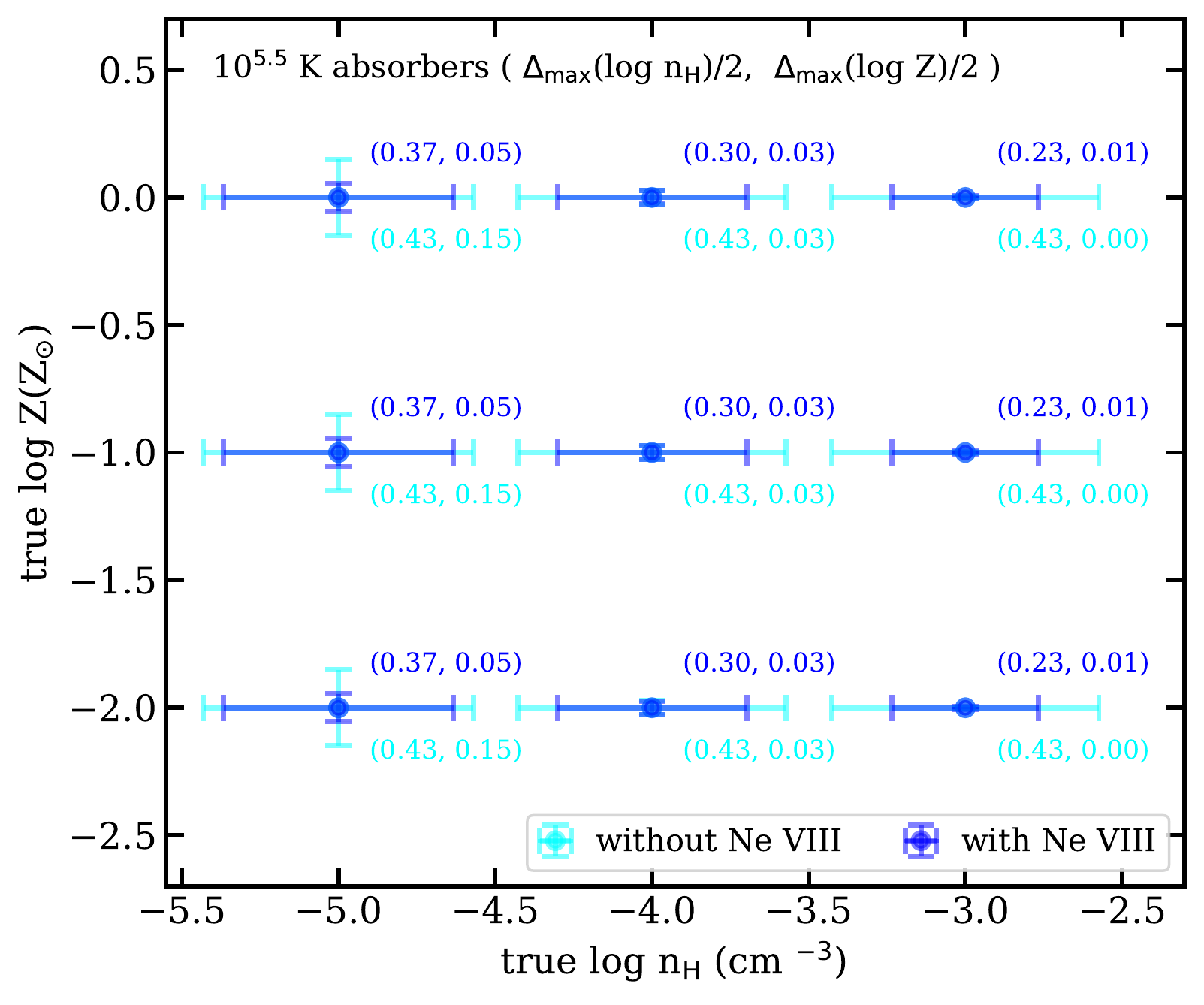}
\caption{The grid of true \nh~and \z~where the extent of error-bars show the 
uncertainty (\dmax~values) on the  inferred \nh~(horizontal error-bars) and  
\z~(vertical error-bars) for warm-hot absorbers at 
$T = 10 ^{5.5}$ K from all nine UV background models. 
The annotated values in the brackets show the \dnmax/2 and \dzmax/2 which can
be quoted as systematic uncertainty on the inferred values of \nh~and \z~using
CGM observations. Blue and cyan error-bars and annotated values indicate 
results when we include 
Ne~{\sc viii} column density for the inference and when we do not, respectively.
The values for former are also given in Table~\ref{tab2}.
}
\label{fig.res_hybrid_all}
\end{figure}

We find that for the same true \nh~and \z~there is a small change in
\dnmax~values of the order of 0.1 dex if we use different true UV 
background models in our mock warm-hot absorbers. Therefore, we repeat the analysis
by changing the true UV background models and then take the mean of
the \dmax~values from all nine true UV background models. Final values 
of \dnmax~and \dzmax~are given in Table~\ref{tab2} and also 
shown in Fig.~\ref{fig.res_hybrid_all} as
error-bars on the true values of \nh~and \z. The blue error-bars show the
extend of \dmax~when we include Ne~{\sc viii} in the ions used for inference
and cyan error-bars when we do not include  Ne~{\sc viii}. 

The annotated values in the figure are the \dnmax/2 and \dzmax/2 which are
equivalent to being systematic uncertainties on the 
log \nh~and log \z~arising from  uncertain UV background. 
When we do not use Ne~{\sc viii} then inferred values of \nh~do not
show any dependence on the true \nh~values and for all the cases the 
\dnmax$\sim 0.86$. In this case, there is more scatter in the inferred \z~at
true \nh$=10^{-5}$ cm$^{-3}$ giving rise to \dzmax$=0.3$ dex as compared
to 0.1 dex when  Ne~{\sc viii} is included in the inference. Whereas,
for true \nh$> 10^{-5}$ cm$^{-3}$, the \dzmax~is negligibly small and
does not depend on inclusion or exclusion of Ne~{\sc viii}.
With increasing true $n_{\rm {H}}$, \dzmax~reduces rapidly 
and approaches to zero at $n_{\rm H} \sim 10^{-3}$ cm$^{-3}$. \emph{Therefore
one can safely say that for warm-hot absorbers at $n_{\rm H} 
\gtrsim 10^{-4}$ cm$^{-3}$ inferred metallicity
is robustly estimated.}
Fig.~\ref{fig.res_hybrid_all} shows that \dnmax~values do not depend on
the true \z, whereas  
\dnmax~changes from 0.74 to 0.46 dex when true \nh~changes from
10$^{-5}$ to 10$^{-3}$ cm$^{-3}$ (when we 
include  Ne~{\sc viii} in the inference). Similar to
the case of photoionized absorber, we
find that both \nh~and \z~are better
constrained for gas at high densities. 

Overall, for the warm-hot absorbers at $T = 10^{5.5}$ K, 
\z~is robustly estimated (within 0.05 dex) 
for gas with density $\ge 10 ^{-4}$ cm$^{-3}$ whereas
the inferred \nh~can be uncertain up to a value of \dnmax~ranging from 0.46 to 0.74 dex for the gas with density 10$^{-3}$ to 10$^{-5}$ cm$^{-3}$. 
However, excluding Ne~{\sc viii} from the analysis can increase the \dnmax~up to 
0.86 dex. Table~\ref{tab2} provides values of \dmax~for 
different true \nh~and \z~for all nine UV background (shown in 
Fig~\ref{fig.res_hybrid_all}) as well as for only seven \citetalias{KS19} UV 
background models.  
However, note that our results assume that observers know the
true temperature of the gas and therefore the quoted uncertainties are only related to
the varying models of UV background.

\subsection{Shape and Normalization of the UV Background}\label{sec.shape_and_norm}
\begin{figure}
\includegraphics[width=0.46\textwidth,height=\textheight,keepaspectratio]{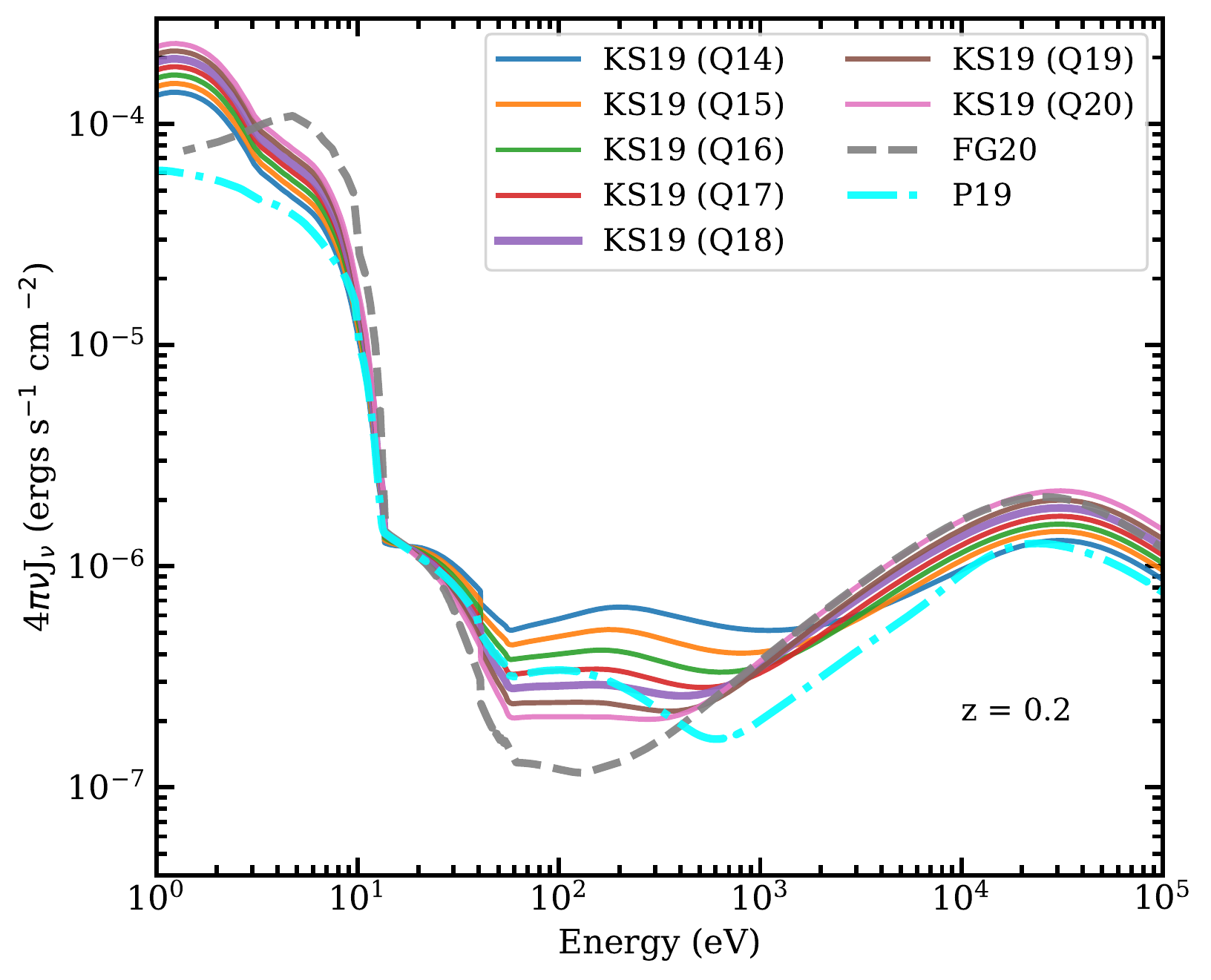}
\caption{
Rescaled UV background models at $z= 0.2$.  Here, 
every UV background spectrum has been normalised so that all of them have same
$\Gamma_{\rm HI}$ as in Q18 model of \citetalias{KS19}. Legends are 
the same as given in Fig.~\ref{fig.uvb}.
}
\label{fig.rescaled_uvb}
\end{figure}

Here, in order to understand the trend in the inferred \nh~and \z~as well as
to find out how much
$\Delta_{\rm max} (\log n_{\rm H})$ and  $\Delta_{\rm max} (\log Z)$ 
is contributed by the change in UV background shapes as compared to its
normalization (or $\Gamma_{\rm HI}$), we normalize all UV background models to have
same $\Gamma_{\rm HI}$ value and repeat the analysis for the test case. 
This normalization is performed by scaling
all UV background models to have same $\Gamma_{\rm HI}$ as 
in Q18 \citetalias{KS19} model. In Fig.~\ref{fig.rescaled_uvb}, we show these 
normalized UV background spectra at $z=0.2$.
Because of the simple scaling to keep $\Gamma_{\rm HI}$ same, all models are close to
each other at $E \sim 13.6$ eV but they differ at high and low energies (compare
Fig.~\ref{fig.uvb} and~\ref{fig.rescaled_uvb}). 
We use these $\Gamma_{\rm HI}$ normalized UV background 
models and repeat our analysis for the test case (shown in 
Fig.~\ref{fig.result_toy_photoionized} and \ref{fig.result_toy_hybrid}) 
for both photoionized and hybrid warm-hot mock absorbers generated 
using Q18 \citetalias{KS19} as a true UV background.

\begin{figure*}
\begin{tabular}{cc}
  \includegraphics[width=0.9\textwidth]{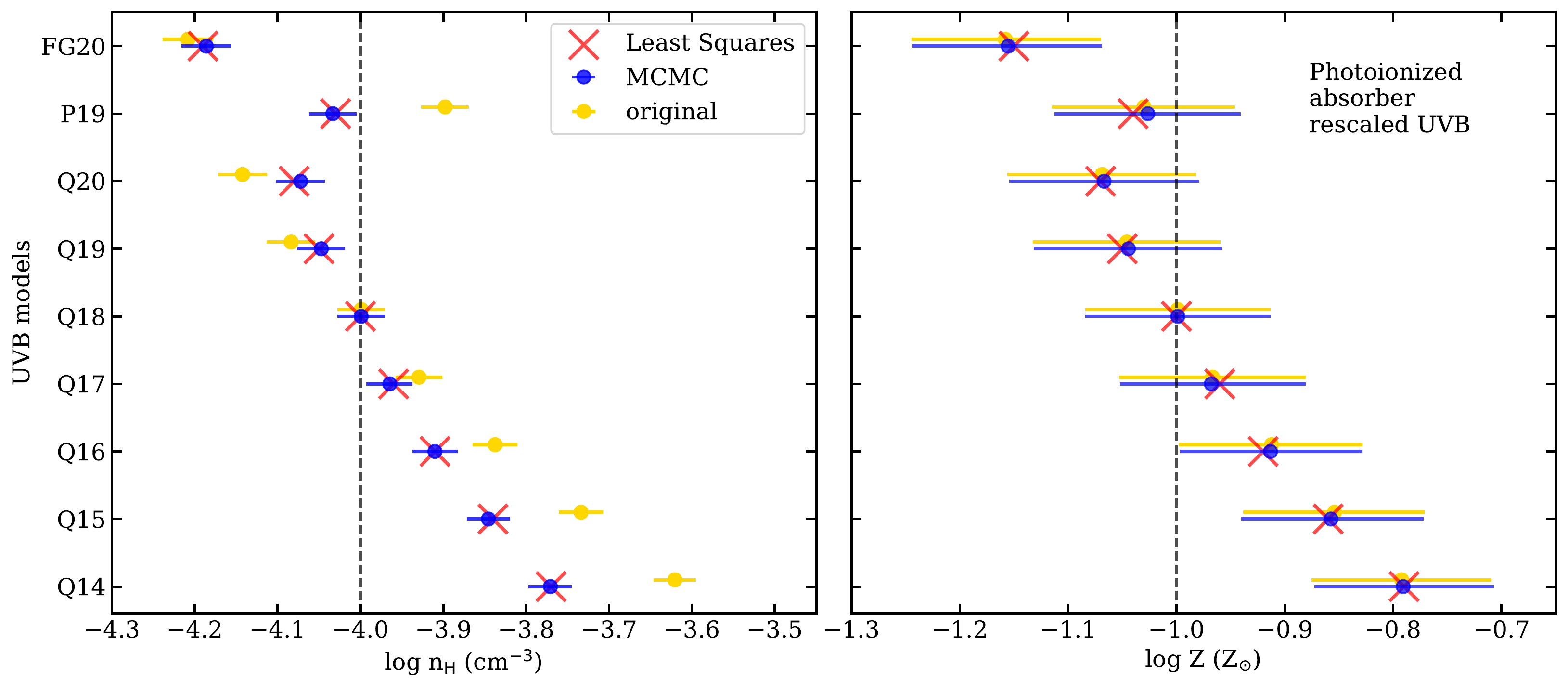}\\
  \includegraphics[width=0.9\textwidth]{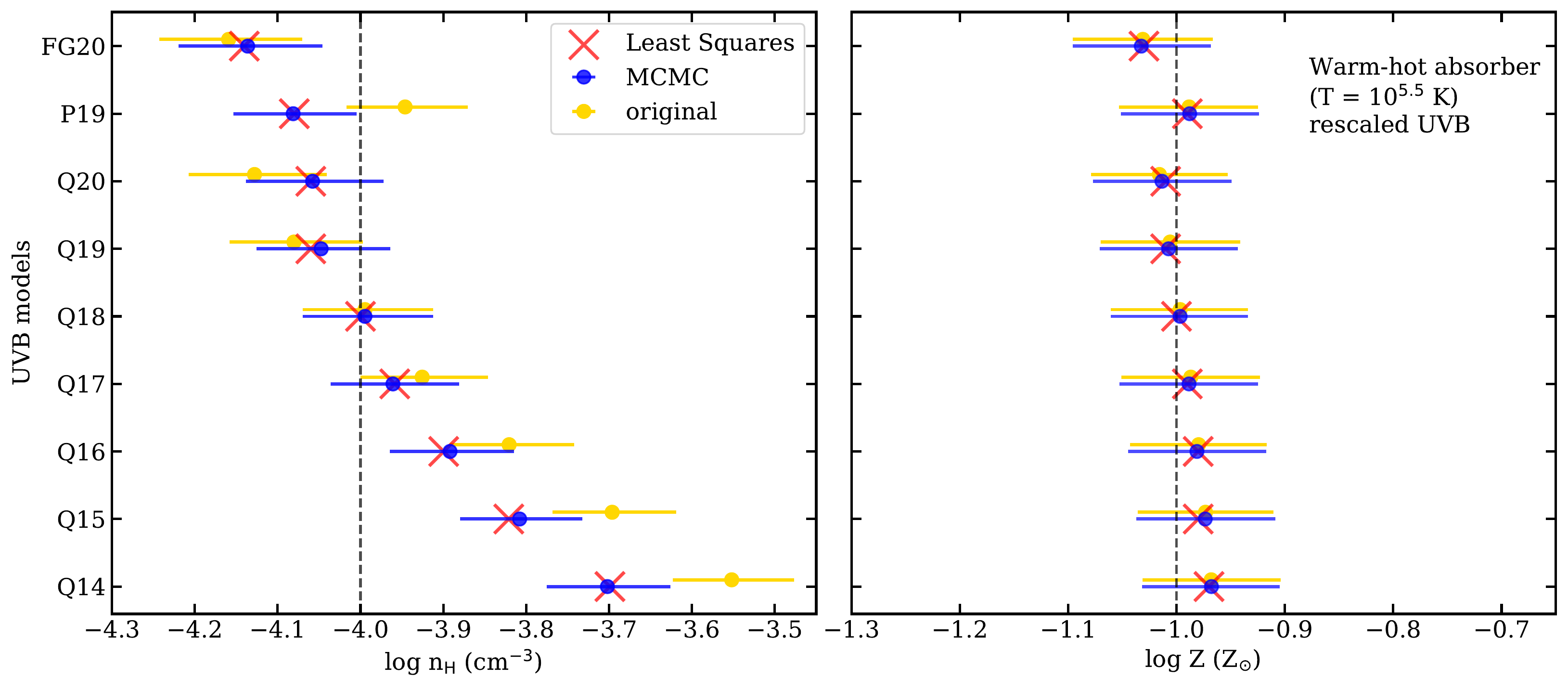} 
\end{tabular}
\caption{Inferred values of $n_{\rm H}$ (cm$^{-3}$; left-hand panel) and $Z$ 
(Z$_{\odot}$; right-hand panel) for re-scaled UV background models 
(labeled on $y-$axis; 
see  Fig.~\ref{fig.rescaled_uvb}) that have same $\Gamma_{\rm HI}$ values.
Results for the photoionized (top panels) and the warm-hot ($T= 10^{5.5}$ K) 
mock observations are shown along with the previous results 
(with yellow data points) obtained for original UV background (presented in Fig.~\ref{fig.result_toy_photoionized} and \ref{fig.result_toy_hybrid}).
Models Q14-Q20 are from \citetalias{KS19}.
The blue circles with horizontal error-bars show the result 
from the Bayesian MCMC method (median with 16 and 84 percentile values of posterior PDF;
see Fig.~\ref{figA} for few illustrations)
while the red crosses shows the result of least square minimization method. 
Vertical dashed lines indicate the true values from toy observations obtained using 
Q18 \citetalias{KS19} UV background  model. 
The same results in metallicity for original and scaled UV background models show that
the metallicity is independent of scaling and depends only on
the shape of the UV background. As expected, for both photoionized and hybrid models, 
the inferred $n_{\rm {H}}$ (blue points) changes by same scaling factors used for
UV background from the previous inferred values (yellow point). 
For photoionized models, total difference in the inferred \nh~arising only from 
change in shape of UV background models $\Delta_{\rm max} (\log n_{\rm H}) = 0.4$ 
and metallicity change is same as original UV background models
$\Delta_{\rm max} (\log Z) =0.66$. For warm-hot absorber
at $T = 10^{5.5}$ K the total variation in density
$\Delta_{\rm max} (\log n_{\rm H}) = 0.44$ as compared to original $0.56$ dex.
}
\label{fig.result_rescale}
\end{figure*}

Results of this exercise are summarized in the Fig~\ref{fig.result_rescale}.
We also show the previous results (from Fig.~\ref{fig.result_toy_hybrid} and 
\ref{fig.result_toy_photoionized}) obtained 
using original UV background models for comparison.
We get the same inferred $Z$ for both photoionized and
hybrid absorbers (right hand panels in Fig~\ref{fig.result_rescale}) as in 
original UV background models. It clearly illustrates that the 
\emph{the main reason behind the variation in the inferred metallicity 
is the change in shape of UV background models whereas 
the scaling of UVB has no effect on the metallicity}. 
Shape of a UV background model 
decides the relative values of photoionization rates for all metal species
which are independent of the scaling of UV background. The scaled UV background
uniformly scales all metal-ion column densities, therefore inferred \z~does not 
change with scaling of UV background. However, in order to reproduce 
these uniformly scaled metal ion column densities, inferred \nh~gets scaled
with same factor as UV background.
We see it in the inferred  $n_{\rm H}$ for
re-scaled UV background models (blue points in Fig.\ref{fig.result_rescale})
as compared to the one with original UV background models (yellow points 
in Fig\ref{fig.result_rescale}) where the ratio of new to original \nh~is equal
to the UV background scaling factor.
Such a scaling of \nh~with UV background arises because 
total hydrogen column density $ N_{\rm H}$ is proportional to the 
$\Gamma_{\rm HI}/ n_{\rm H}$ for low density IGM and CGM 
clouds.\footnote{It can be  understood in terms of photoionization 
equilibrium for hydrogen gas where $\Gamma_{\rm HI} \propto 
(1- x_{\rm HI})^2 n_{\rm H} N_{\rm H}$ ignoring small temperature
dependence of hydrogen recombination coefficient  for a fixed value of 
$N_{\rm HI}$. For the low density IGM and CGM gas, the neutral fraction 
$x_{\rm H I} = n_{\rm HI}/ n_{\rm H} << 1$ implies $N_{\rm H} \propto 
\Gamma_{\rm HI}/ n_{\rm H}$.} 
Since inferred \z~ is independent of UV background scaling, 
to obtain the observed column densities of metal ions, $N_{\rm H}$ should be same.
Therefore, $n_{\rm H}$ scales with the $\Gamma_{\rm HI}$ i.e, with
the scaling factor of the UV background.
 In other words, this scaling of \nh~is a result  of preserving ionization parameter for all ions since scaling UV background uniformly scales $n_{\gamma}$
by same amount for all metal ions. Therefore, to reproduce abundance pattern of all ions
the inferred \nh~scales with the same UV background scaling.

For such scaled UV backgrounds, we see  
reduced variation in the inferred $n_{\rm H}$ values for both 
photoionized and warm-hot absorbers as compared to the values 
obtained with original UV background models.
In case of photoionized models, 
we find $\Delta_{\rm max} (\log n_{\rm H}) = 0.4 $ dex
as compared to previous (with un-normalised original UV background) 
value of 0.66 dex, and in case of warm-hot gas the 
$\Delta_{\rm max} (\log n_{\rm H}) = 0.44 $ dex
as compared to previous value of 0.56 dex.
The still significant values of $\Delta_{\rm max} (\log n_{\rm H}) \sim 0.4$ dex
despite the fact that $\Gamma_{\rm HI}$ is same for all UV background models,
suggest that \textit{the inferred \nh~depends both on the normalization and shape of the UV
background}. Here, the variation \dnmax$\sim 0.4$ dex is arising due to change in 
shape of the UV background models. This is mainly due to jointly 
constraining the metallicity and density. 
The relative change in metal column densities arising due to shape of UV background
demands different density of the gas. In order to
naively understand the trend, consider following argument; for a given value of 
$n_{\rm H}$ a harder (softer) UV background will result into lower (higher) 
column densities of metal ions 
therefore the inference will prefer  higher (lower) density of gas than $n_{\rm H}$
so there is more (less) self shielding to match (higher) lower 
values of observed metal column densities. Note that, the variation \dnmax~is too large to be explained with only ionization parameter of hydrogen.

To understand the trend in inferred \nh~and \z, it is better to have UV background
models that vary minimum number of parameters, as in the case of seven models of
\citetalias{KS19}. For example, even though the \citetalias{Puchwein19} model 
use $\alpha = 1.7$ as quasar SED, the UV background spectrum is not same as Q17 
model of \citetalias{KS19} or \citetalias{FG20} which also use $\alpha = 1.7$. 
It is because other parameters such as the quasar emissivity at $13.6$ eV or the
distribution of H~{\sc i} gas in IGM are different, which changes the shape of 
the UV background spectrum. We find that the spectrum of \citetalias{Puchwein19} 
at E $<50$ eV lies somewhere in between Q18 and Q19 models. This explains trend 
in the inferred metallicity for \citetalias{Puchwein19} UV background
(e.g., in Fig.~\ref{fig.photo_all_1}, \ref{fig.photo_all_2} and \ref{fig.hybrid_all_2}). 
The $\Gamma_{\rm HI}$  for \citetalias{Puchwein19} is higher than the Q18 model 
(at $z= 0.2$; see Fig.~\ref{fig.gamma_HI}) which results into higher inferred 
\nh~than for Q18 (e.g., in Fig.~\ref{fig.photo_all_1}, \ref{fig.photo_all_2}).
Note that the inferred \nh~does depend on both spectrum and the normalization,
therefore when we normalize \citetalias{Puchwein19} to same $\Gamma_{\rm HI}$ of 
Q18 model, we infer the lower \nh~values, as shown in Fig.~\ref{fig.result_rescale}.

\subsection{Other Uncertainties and Caveats}\label{sec:4.4}
We recommend quoting $\pm$\dnmax/2 and $\pm$\dzmax/2 
from Fig.~\ref{fig.FigB} for clouds with different $N_{\rm HI}$ 
and Fig.!\ref{fig.res_hybrid_all} for hybrid absorbers 
as a systematic uncertainty on the inferred \nh~and \z~arising from 
uncertain UV background models. For example, in case of $N_{\rm HI}\sim 10^15$ 
cm$^{-2}$ absorber if the inferred 
log \nh(cm$^{-3}$) and log\z/$Z_{\odot}$ from some CGM observations
assuming photoionization equilibrium results in to values of
$-4.8$ and $-1.2$, one can read off the \dnmax~and \dzmax~from its closest
point (log\nh, log\z)$\equiv(-5, -1)$ in Fig~\ref{fig.res_photo_final} 
(or Table~\ref{tab1})
and quote $\pm 0.4$ and $\pm 0.23$ as a systematic uncertainty on 
log \nh~and log \z, respectively. We checked the redshift evolution of the \dnmax~and
\dzmax~for both photoionized and warm-hot absorber and found no
significant change in the values at $z<1.5$. 

Although we recommend using $\pm$\dnmax/2 and $\pm$\dzmax/2
as a systematic uncertainty, we caution readers
that it need not be symmetric. For example if one uses \citetalias{FG20}
background model to infer the 
\nh~then the uncertainty should be $+$\dnmax~and $-0$, given
the fact that all other models, being harder that \citetalias{FG20}, infer
higher \nh. Moreover, note that the uncertainties obtained here are under the 
assumption of simplest possible CGM absorbers with uniform density and
metallicity. Complex multi-phase configuration with non-equilibrium conditions 
may require their own analysis.
However note that, as long as a part of cloud assumed to have uniform density 
and metallicity irradiated with a UV background, uncertainties arising from 
UV background will be similar to the one presented here. For example, even for
multi-phase gas as invoked in many studies 
\citep[e.g,][]{ Pachat16, Pachat17, Nathegi21, Haislmaier21}, each phase 
will have similar uncertainty from UV background. Even with innovative techniques
to models individual cloudlets as presented in \citet{Sameer21}, each cloudlet and
its phase will suffer with same uncertainty from UV background as long 
as a preferred UV background model is fixed for the inference.  

In real observations, fitting uncertainties on the \nh~and 
\z~(say $\Delta_{\rm fit}$) can be obtained
from various fitting methods used to find solution for \nh~and \z,
such as the confidence intervals from our Bayesian MCMC approach.
These uncertainties depend on the column density measurement errors of ions used 
for the inference. 
In the calculations presented here,  we assumed a fixed value of $N_{\rm HI}$ 
and used it as a stopping criteria for cloudy models, whereas observations will
have measurement errors $\Delta N_{\rm HI}$, for example, from 
Voigt-profile fitting of H~{\sc i} absorption lines. 
To find out how much  $\Delta N_{\rm HI}$  contributes to the error budget
on inferred \nh~and \z~we ran cloudy models with different $N_{\rm HI}$
as a stopping criteria and found that the inferred \nh~is independent of 
$N_{\rm HI}$ values but the inferred \z~scales linearly with the 
change in $N_{\rm HI}$. This is the reason for the correlation seen in the 
2D posteriors for photoionized gas 
(as in Fig.~\ref{figA} and right-hand side of Fig.~\ref{fig.inference}).
Therefore,
total error on \nh~has to be the sum of systematic error from UV background models
(\dnmax/2) and  $\Delta_{\rm fit}$ (i.e here, confidence intervals from 
Bayesian MCMC fits). Whereas, error on inferred \z~should include
\dzmax/2, $\Delta_{\rm fit}$ and $\Delta N_{\rm HI}$. Therefore, 
as long as the $\Delta N_{\rm HI} >>\,$\dzmax/2, systematic uncertainty 
from UV background may not play an important role in total error budget of 
inferred \z. 
Note that we are only discussing the errors on \nh~and \z~arising from UV background
and their relation to the observed errors on $N_{\rm HI}$. There are other potential
effects such as poorly known gas temperature that will have its own 
uncertainty, quantifying which is beyond the scope of this paper.

\begin{table}
\centering
\caption{Maximum variation in $n_{\rm H}$ and $Z$ for photoionized absorber 
with $N_{\rm HI} = 10^{15}$ cm$^2$
(from Fig.~\ref{fig.res_photo_final}), refer to 
Fig.~\ref{fig.FigB} for different $N_{\rm HI}$.}
\begin{tabular}{|c|c|c|c|c|c|}
\hline
 True & True & \dmax & \dmax & \dmax & \dmax \\
 log \nh & log \z & (log \nh) & (log \z) & (log \nh) & (log \z) \\
 (cm $^{-3}$) & ($Z_{\odot}$) & all UVB &  all UVB & KS19  &  KS19 \\
 \hline
 -5 & -2 & 0.79 & 0.44 & 0.69 & 0.36\\
-5 & -1 & 0.80 & 0.47 & 0.69 & 0.36\\
-5 & 0 & 0.75 & 0.57 & 0.67 & 0.42\\
-4 & -2 & 0.72 & 0.57 & 0.61 & 0.41\\
-4 & -1 & 0.72 & 0.57 & 0.60 & 0.41\\
-4 & 0 & 0.65 & 0.56 & 0.55 & 0.40\\
-3 & -2 & 0.54 & 0.35 & 0.49 & 0.24\\
-3 & -1 & 0.54 & 0.35 & 0.49 & 0.24\\
-3 & 0 & 0.52 & 0.36 & 0.47 & 0.25\\
\hline

\end{tabular}
\begin{flushleft}
\footnotesize{Note: Here, `all UVB' refers to all nine models of UV background while `KS19' refers to seven models of \citetalias{KS19}.} \\
\end{flushleft}
\label{tab1}
\end{table}

\begin{table}
\centering
\caption{Maximum variation in $n_{\rm H}$ and $Z$ for warm hot absorber 
at $T = 10^{5.5}$ K  including Ne~{\sc viii} (see Fig.~\ref{fig.res_hybrid_all})}
\begin{tabular}{|c|c|c|c|c|c|}
\hline
 True & True & \dmax & \dmax & \dmax & \dmax \\
 log \nh & log \z & (log \nh) & (log \z) & (log \nh) & (log \z) \\
 (cm $^{-3}$) & ($Z_{\odot}$) & all UVB &  all UVB & KS19  &  KS19 \\
\hline
-5 & -2 & 0.73 & 0.11 & 0.67 & 0.05\\
-5 & -1 & 0.73 & 0.11 & 0.67 & 0.05\\
-5 & 0 & 0.73 & 0.11 & 0.67 & 0.05\\
-4 & -2 & 0.60 & 0.05 & 0.57 & 0.04\\
-4 & -1 & 0.60 & 0.05 & 0.57 & 0.04\\
-4 & 0 & 0.60 & 0.05 & 0.57 & 0.04\\
-3 & -2 & 0.47 & 0.01 & 0.47 & 0.01\\
-3 & -1 & 0.47 & 0.01 & 0.47 & 0.01\\
-3 & 0 & 0.47 & 0.01 & 0.47 & 0.01\\
\hline

\end{tabular}
\begin{flushleft}
\footnotesize{Note: Here, `all UVB' refers to all nine models of UV background while `KS19' refers to seven models of \citetalias{KS19}.} \\
\end{flushleft}
\label{tab2}
\end{table}
.

\section{Summary and conclusion }\label{sec:5}

We studied the effect of using different UV background models on the 
inferred density $n_{\rm H}$ and metallicity $Z$ of CGM absorbers. 
We chose nine viable UV background models from most recent and updated 
studies (see Fig.~\ref{fig.uvb}) which include seven models generated 
with varying quasar SED from \citetalias{KS19} and two models from \citetalias{FG20} and \citetalias{Puchwein19}.

First, using {\sc cloudy} software for ionization modelling, we created 
two sets of toy models one for photoionized absorber and other for 
collisionally ionized warm-hot absorber. 
We irradiated these absorbers with a fiducial UV background model and 
noted down the model column densities of many metal ions commonly 
observed in the CGM. We treat a set of those metal-ion column densities 
along with $N_{\rm HI}$ as our toy observations. Then we use these toy 
observations to infer the $n_{\rm H}$ and $Z$ using different UV 
background models. 

We use two methods to jointly infer the $n_{\rm H}$ and $Z$ from 
such toy observations, the maximum likelihood estimate using the 
Bayesian MCMC method and by minimizing the least square difference in 
the model and observed (from toy models) column densities. 
We show that both methods 
provide identical results. To quantify the variation in the 
$n_{\rm H}$ and $Z$, we define total variation 
$\Delta_{\rm max} (\log n_{\rm H})$  and $\Delta_{\rm max} (\log Z)$ 
which is the maximum difference 
in the inferred median values of $n_{\rm H}$ and $Z$ obtained using
large set of mock observations assuming all nine UV
background models.

For photoionized CGM absorbers, we find that the \dnmax~ranges from 0.5
to 0.8 dex (a factor of 3.2 to 6.3 difference) whereas \dzmax~ranges from
0.2 to 0.6 dex (a factor of 1.6 to 4) difference for CGM clouds 
having density 10$^{-3}$ to 10$^{-5}$ cm$^{-3}$ and hydrogen column $N_{\rm HI}$ ranging 
from  $10^{14}$ to $10^{19}$ cm$^{-2}$. The \nh~and \z~ dependent 
values of \dmax~are shown in Fig.~\ref{fig.FigB}. 

For collisionally ionized CGM absorbers at $T=10^{5.5}$ K, we find that
including Ne~{\sc viii} column density in the inference provides better 
results that are closer to the true values (Fig.~\ref{fig.photo_all_1}).
Interestingly, we find that 
the inferred $Z$ is always robustly estimated (within $\sim 0.1$ dex). 
However, \dnmax~ranges from 0.47 to 0.73 dex (a factor of 3 to 
5.4) difference for CGM clouds having density 10$^{-3}$ to 10$^{-5}$ 
cm$^{-3}$. If Ne~{\sc viii} is not included in the inference, 
the \dnmax~and \dzmax~further increases by $\sim~0.2$ dex. 
The \nh~and \z~ dependent values of \dmax~with and without Ne~{\sc viii} 
for warm-hot absorber are shown in Fig.\ref{fig.res_hybrid_all}
and tabulated in Table~\ref{tab2}. 

We recommend quoting $\pm$\dnmax/2 and $\pm$\dzmax/2 from
Fig.~\ref{fig.FigB}
and Fig.~\ref{fig.res_hybrid_all}
as a systematic uncertainty on the inferred \nh~and \z~arising from 
uncertain UV background models for photoionized and and warm-hot 
absorber, respectively. For a case of $N_{\rm HI} =10^{15}$ cm $^{-2}$ 
absorber, we also provide these \dmax~values in 
Table~\ref{tab1} and \ref{tab2} along with the one obtained only for seven 
UV background models of \citepalias{KS19} where different UV background
were obtained by changing only one parameter i.e shape of quasar SED. 
Note that these results are valid 
for the IGM and CGM absorbers having
density in the range $n_{\rm H } < 10^ {-2}$ cm$^{-3}$. At very high
densities ionization fraction for hydrogen and metals may drop sharply
due to self shielding and then the inferred values may become 
less sensitive to the incident UV background. 

To study which factor, the shape or the amplitude (normalization) 
of the UV background, is important for the obtained variation in the 
$n_{\rm H}$  and $Z$, we ran more {\sc cloudy} models with modified UV 
background by scaling all nine UV backgrounds to have same H~{\sc i} 
photoionization rates ($\Gamma_{\rm HI}$) as in fiducial Q18 
\citetalias{KS19} UV background model.
These re-scaled backgrounds (see Fig~\ref{fig.rescaled_uvb}) 
differ only in the relative shapes. From these models we find that the 
inferred metallicity for photoionized absorbers only depends on the shape 
of the UV background and it is independent of the normalization, whereas
inferred $n_H$ depends on both the shape and 
normalization of UV background. We find,
\emph{harder (softer) UV background implies high (low) \nh~and \z}.

The inferred density and metallicity is directly related to the line-of-sight thickness of CGM clouds and their mass. Thus, the large variation in these parameters suggests that we need to be cautious while interpreting the results of CGM studies and ionization models using just one UV background model. On the other hand, this demands more constraining observations of input parameters used in synthesis of UV background models, especially the ones that determine the shape of the UV background, such as the intrinsic SED of quasars at ionizing energies. This would allow the future UV background models to provide a more precise picture of the CGM.

\section*{acknowledgements} 

AA thanks Robert Antonucci for hosting him at the University of California, Santa Barbara and helpful discussions on the topic. AA acknowledges financial support from the Kishore Vaigyanik Protsahan Yojana (KVPY) Fellowship for undergraduates by the Department of Science and Technology, Government of India. VK is supported through the INSPIRE Faculty Award (No. DST/INSPIRE/04/2019/001580) of the Department of Science and Technology (DST), India. Authors would also like to thank the discussions and suggestions in the KITP Halo21 workshop, supported  in part by the National Science Foundation under Grant No. NSF PHY-1748958. Authors would also like to thank Abhisek Mohapatra for his feedback, suggestions and comments on the paper. Lastly, the authors would like to thank the anonymous referee for their suggestions and comments which helped in improving this paper significantly.

\section*{Data Availability}

No observational data was used for this project. The outputs for both the photoionized and warm-hot absorber models for all UV backgrounds can be made publicly available on request.

\bibliographystyle{mnras}
\bibliography{vikrambib}

\appendix
\section{Posterior PDFs for few example cases}\label{A}
\begin{figure*}
\begin{tabular}{cc}
  \includegraphics[width=85mm]{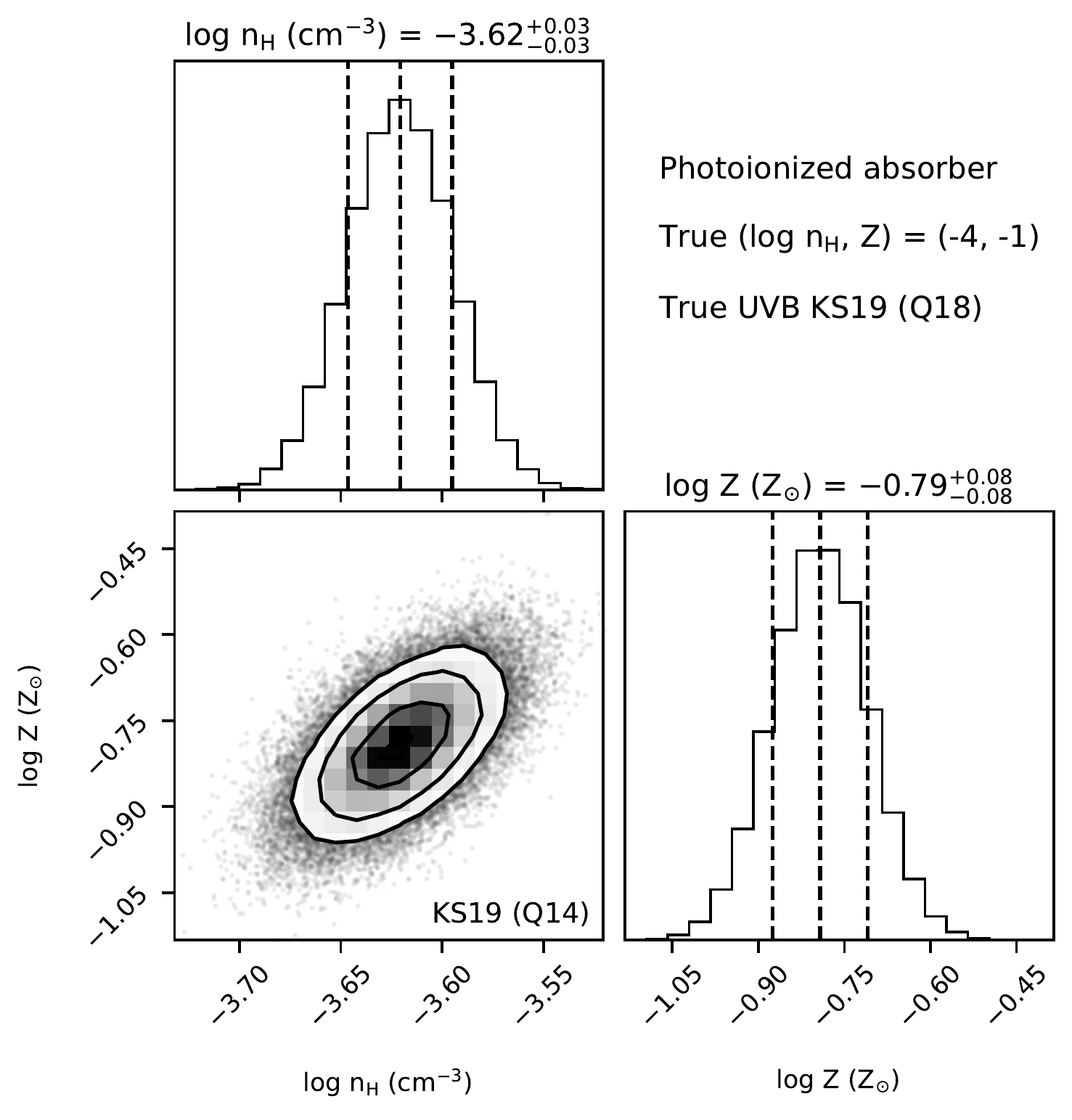} &   \includegraphics[width=86mm]{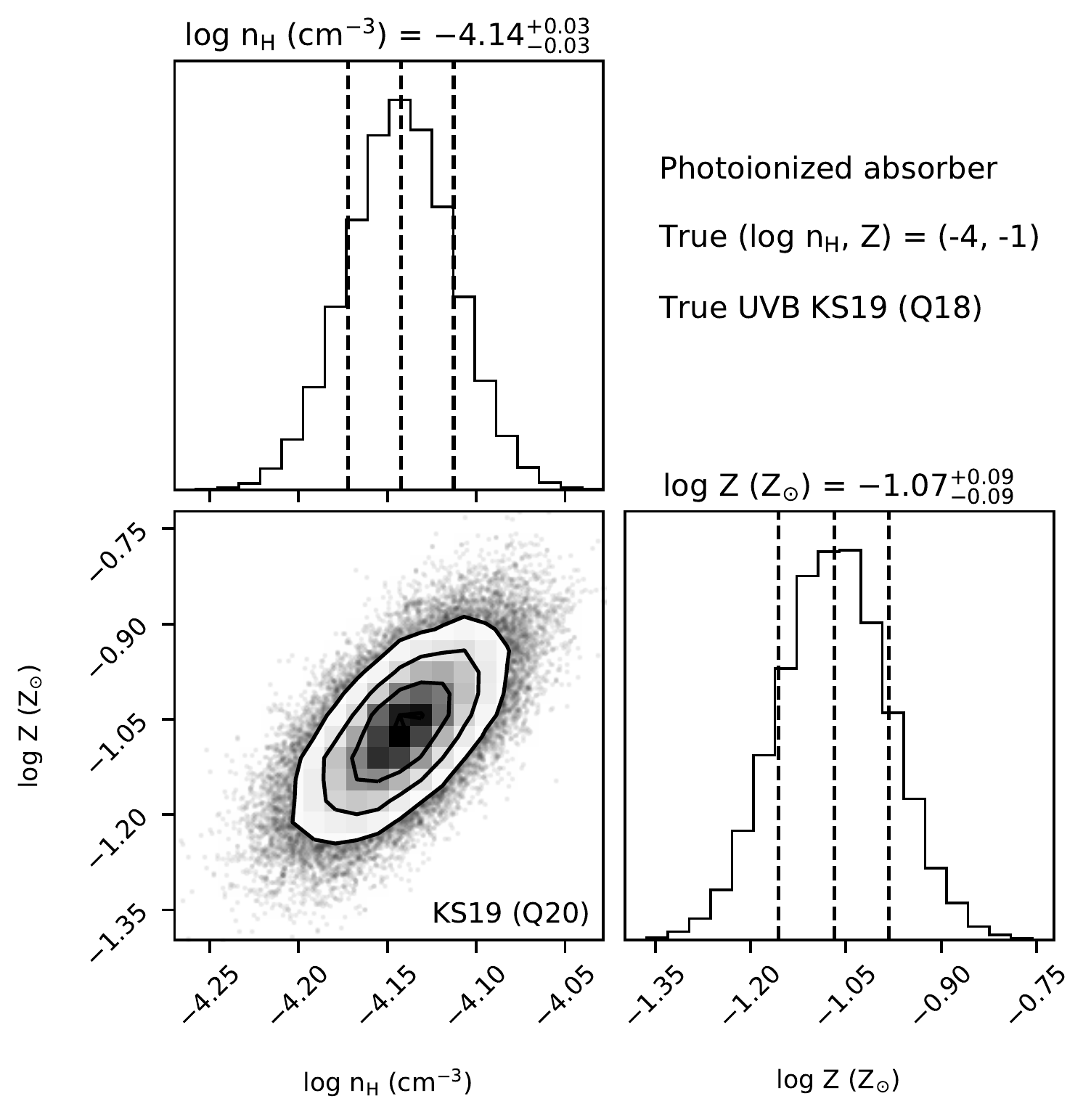} \\
(a) With Q14 \citetalias{KS19} UV background & (b) With Q20 \citetalias{KS19} UV background \\[10pt]
 \includegraphics[width=85mm]{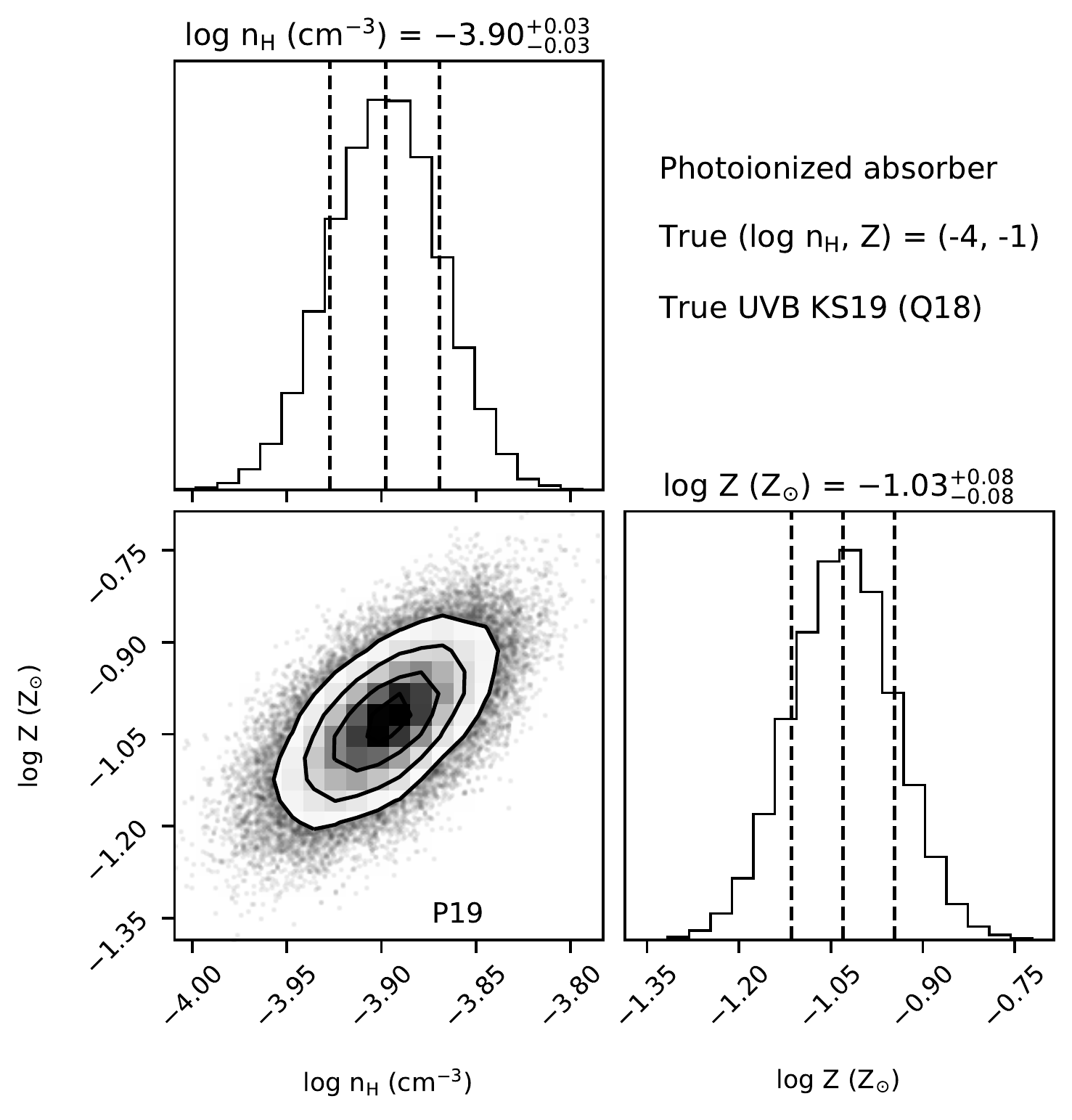} &   \includegraphics[width=86mm]{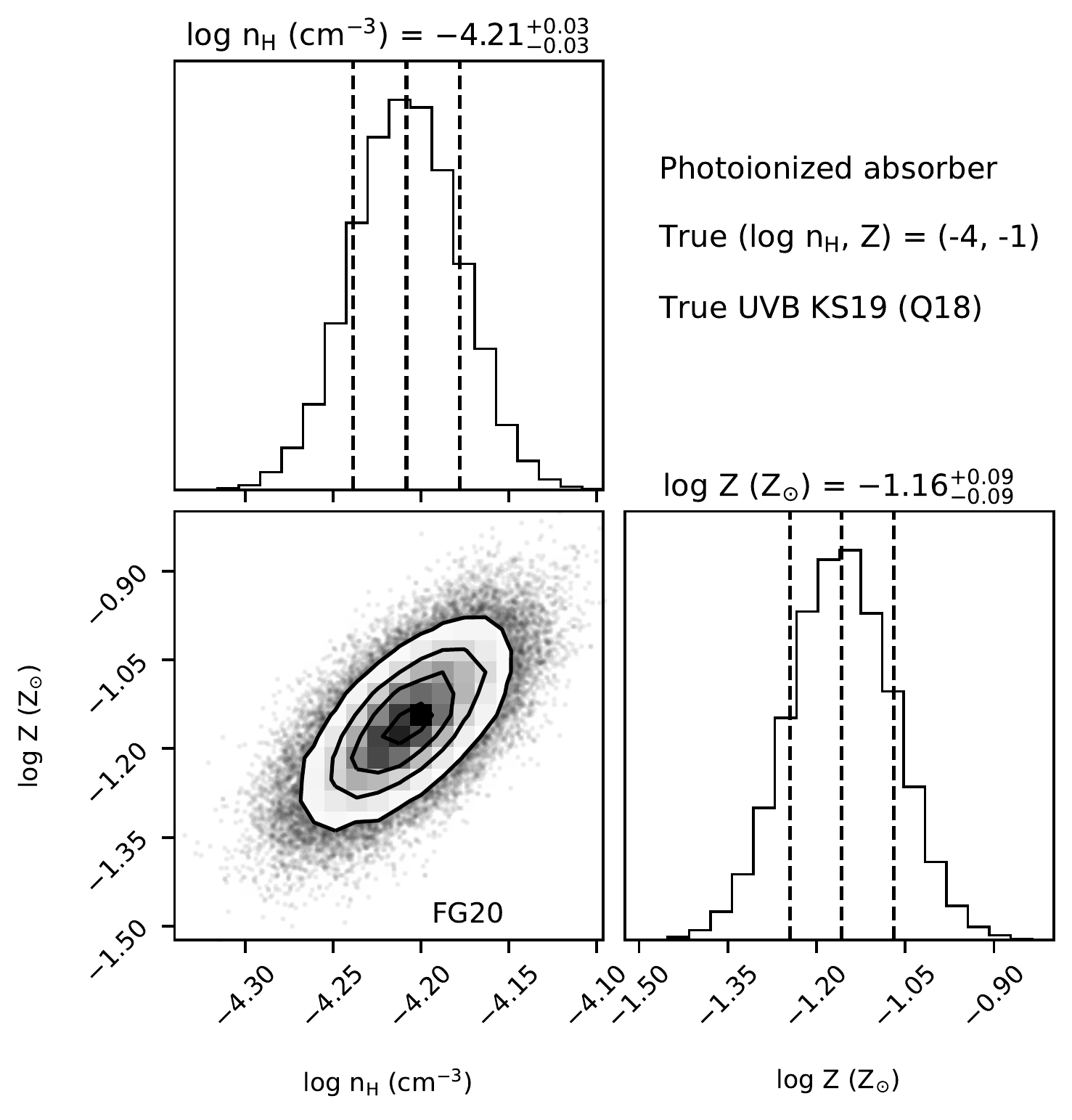} \\
 (c) With \citetalias{Puchwein19} UV background & (d) With \citetalias{FG20} UV background \\[10pt]
\end{tabular}
\caption{Some example posterior PDFs similar to the one shown in Fig. \ref{fig.inference} but the
inference is performed assuming different models. The panel (a) and (b) are obtained 
by using two Q14 and Q20 models of the UV background while panel (c) and (d) are obtained by using
P19 and FG20 models of the UV background. Note that the mock observations are obtained by using
Q18 KS19 UV background models having true values $\log n_{\rm H} (\rm cm^{-3}) = -4$ and 
$\log Z (Z_{\rm \odot}) = -1$.}
\label{figA}
\end{figure*}

\begin{figure*}
\begin{tabular}{cc}
  \includegraphics[width=85mm]{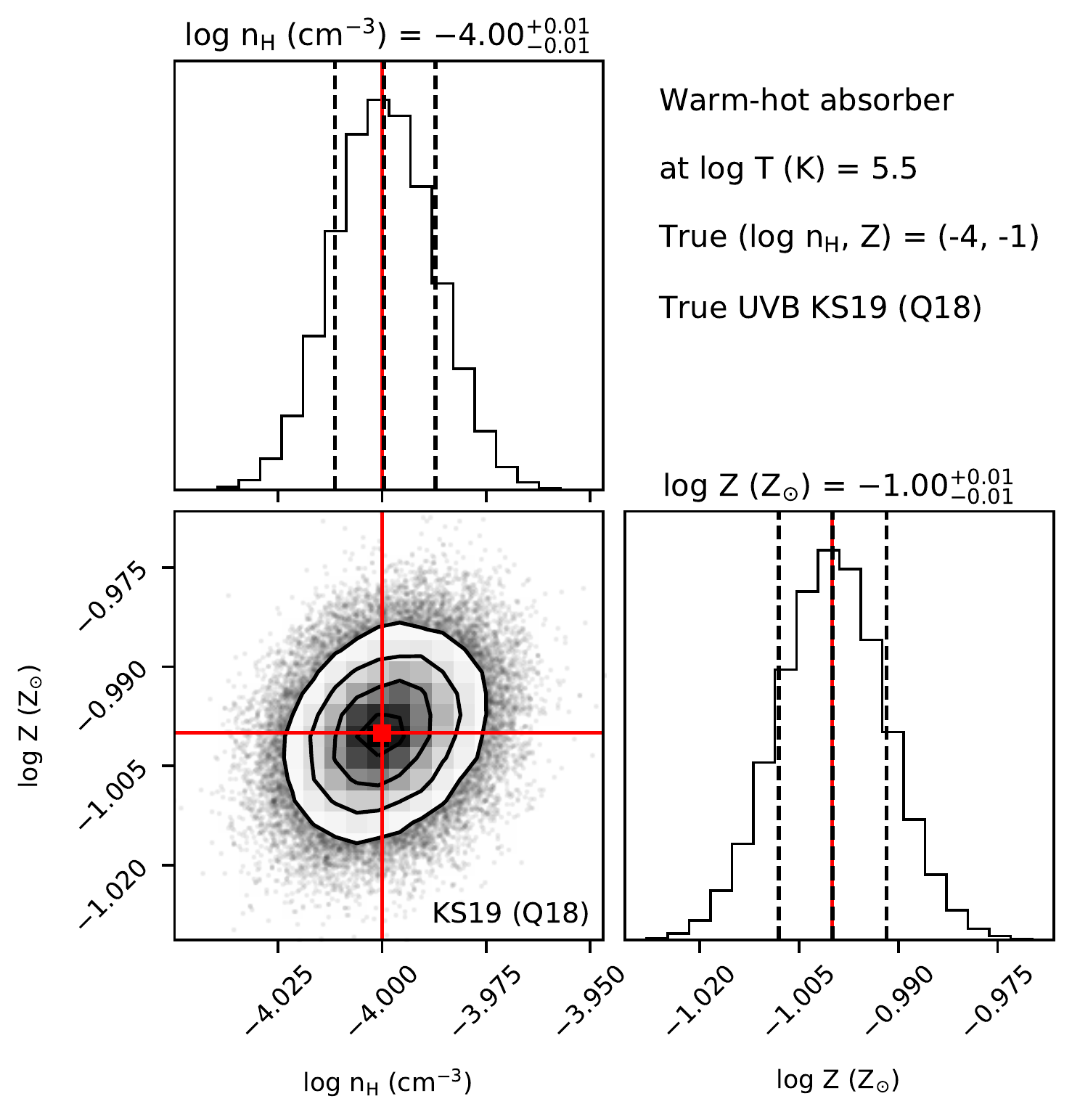} &   \includegraphics[width=86mm]{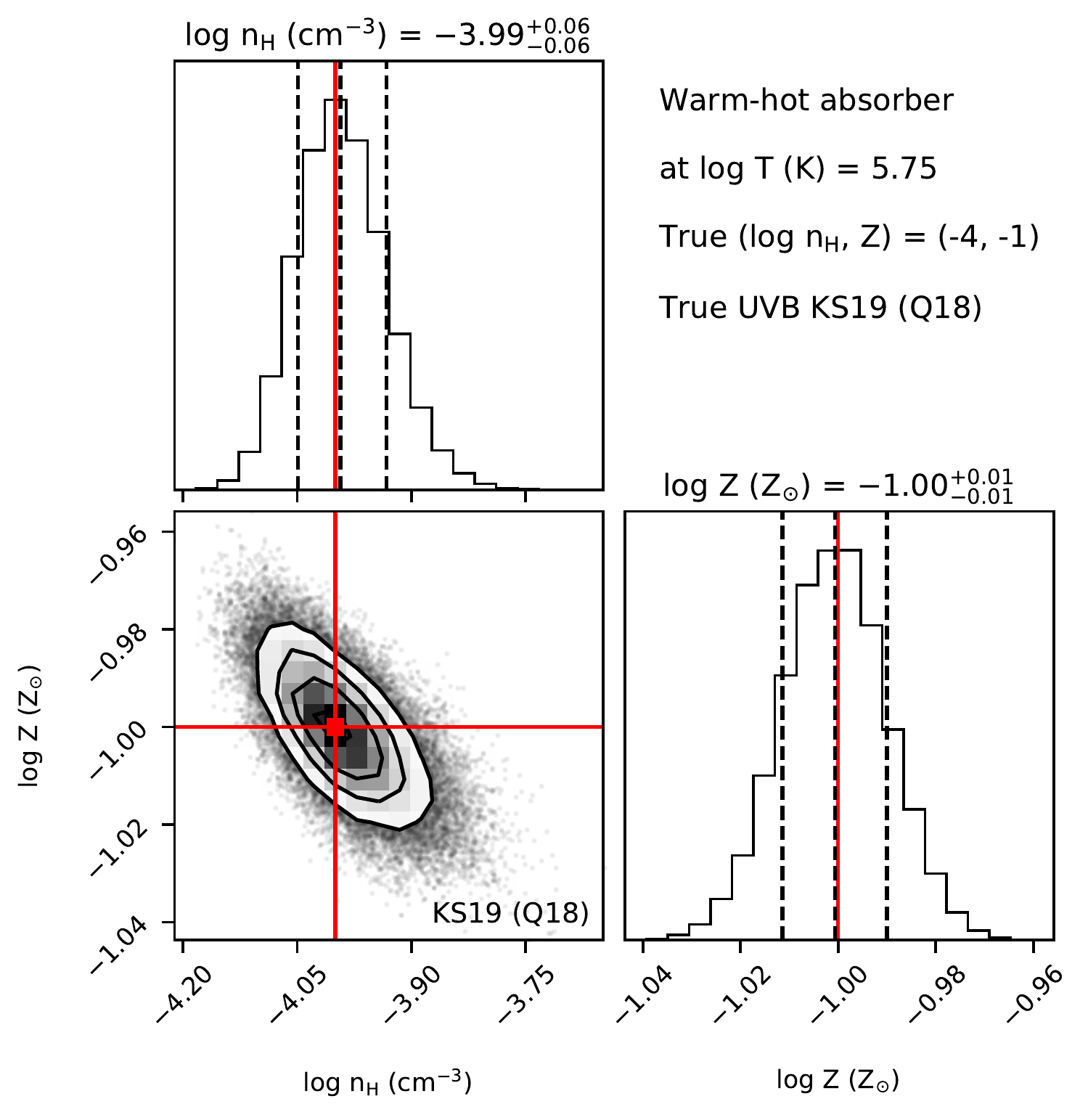} \\
(a) at $T = 10^{5.5} K$ & (b) at $T = 10^{5.75} K$ \\[10pt]
 \includegraphics[width=85mm]{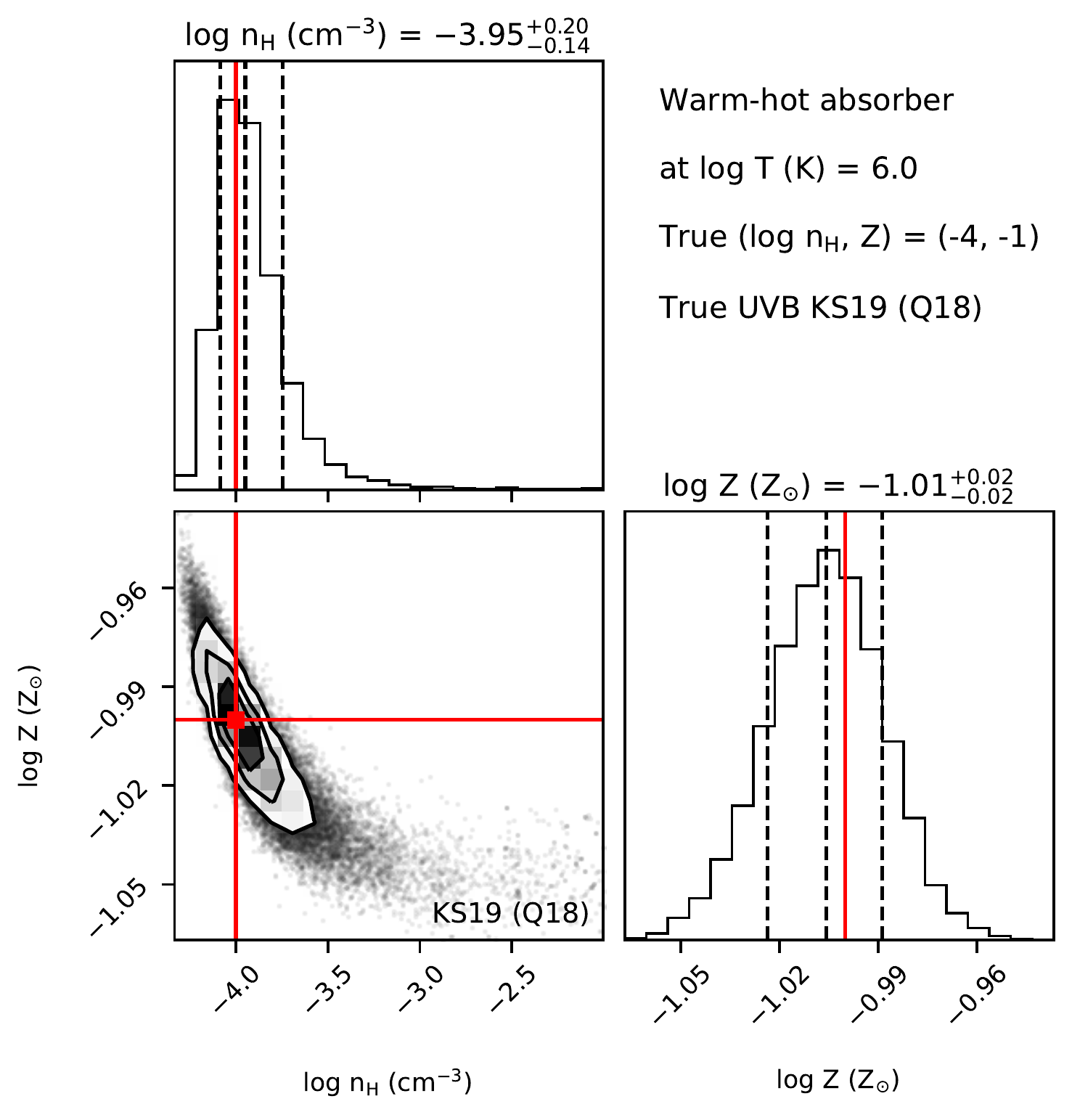} &   \includegraphics[width=86mm]{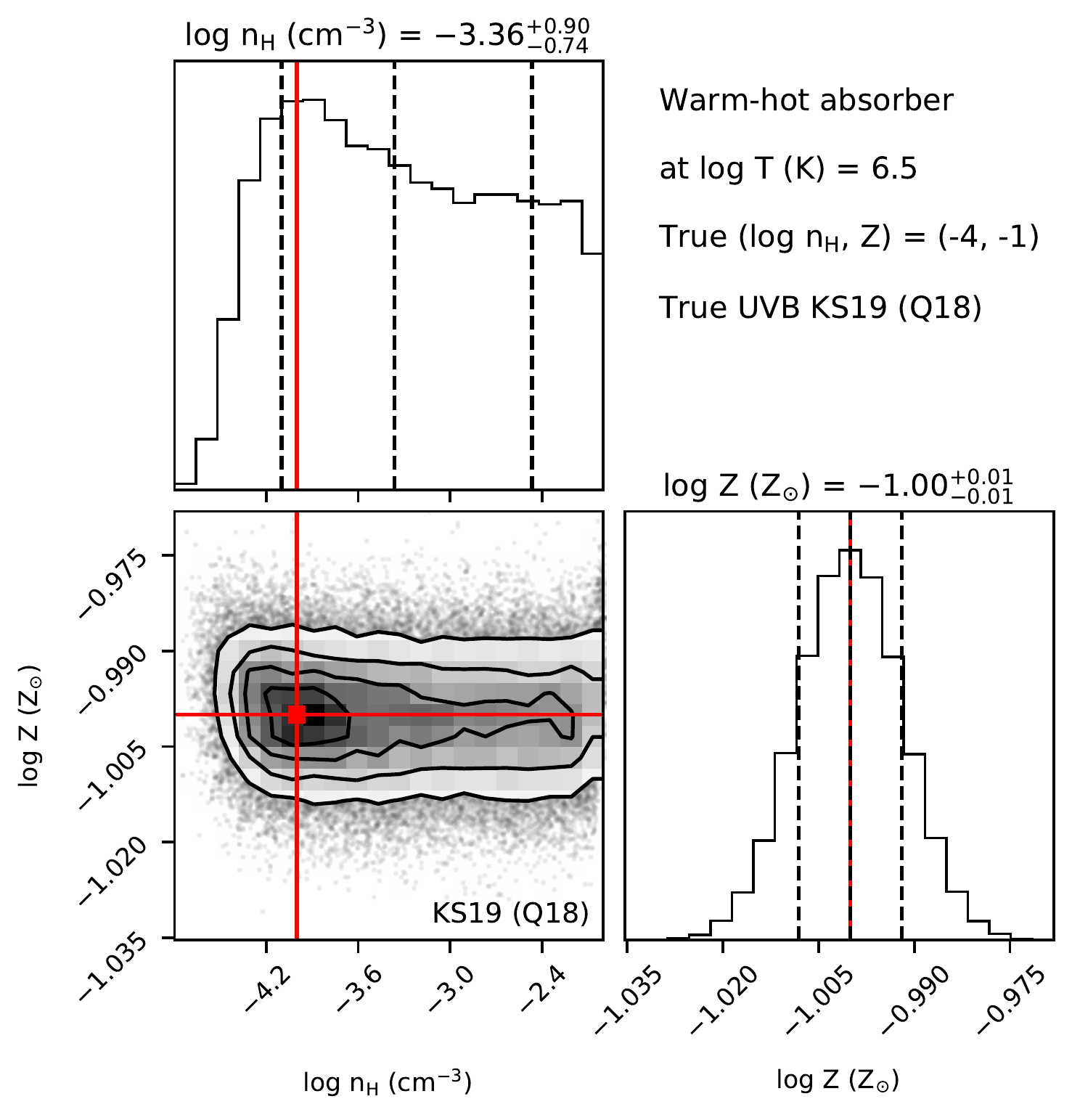} \\
 (c) at $T = 10^{6} K$ & (d) at $T = 10^{6.5} K$ \\[10pt]
\end{tabular}
\caption{Example inference tests for warm-hot absorbers at different 
temperature where red lines indicate the true values. 
For these tests, we used very small errors on the column densities of 
metal ions
($<0.1$ dex) so that posteriors are well behaved upto 
$T = 10^6$K. As we increase temperature, density becomes more 
unconstrained, however metallicity always remains well constrained. }
\label{figB}
\end{figure*}

\section{Results on varying Optical Thickness of Cloud}\label{A}

\begin{figure*}
\begin{tabular}{c|c}
\includegraphics[width=0.5\textwidth,height=\textheight,keepaspectratio]{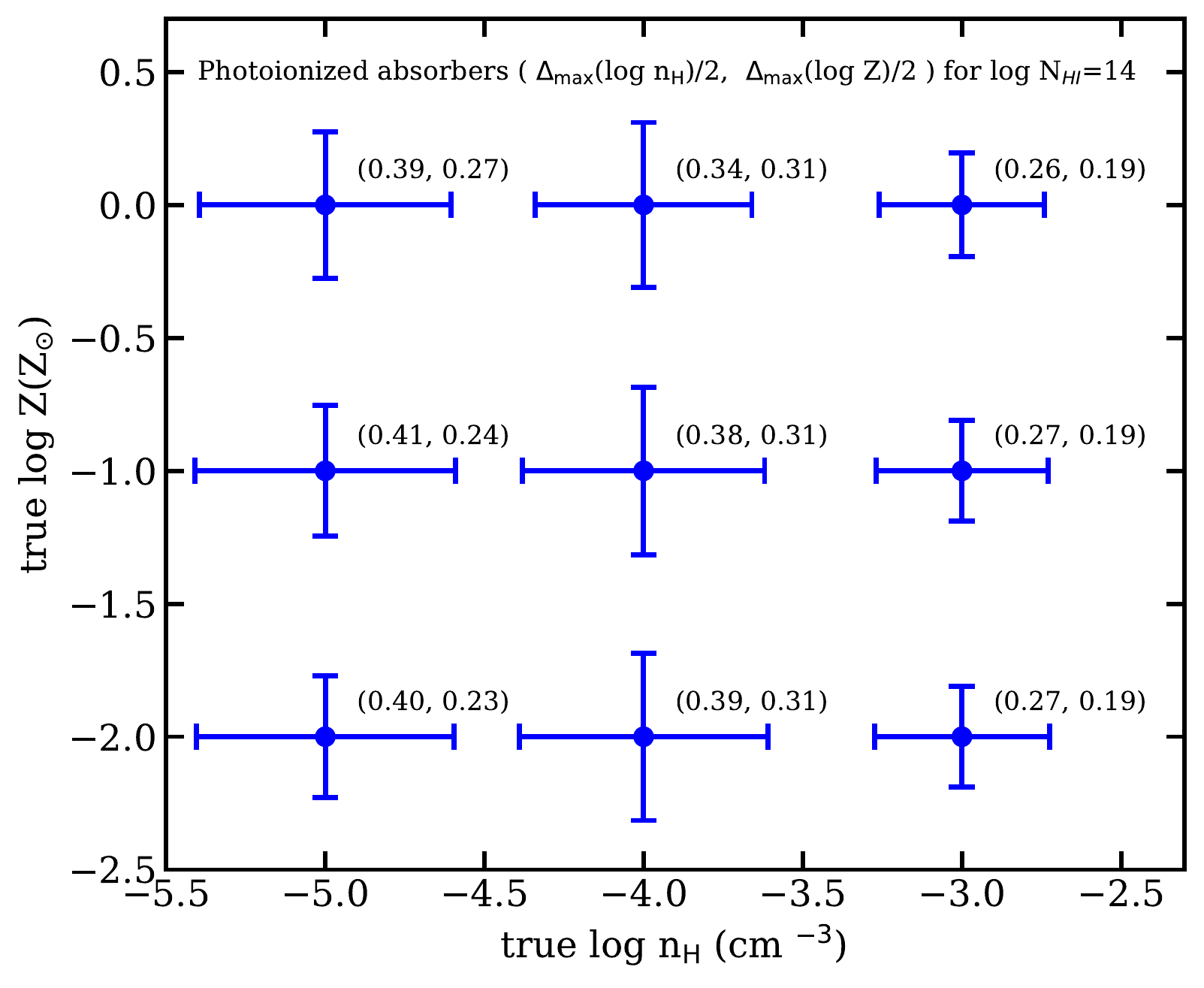} &
\includegraphics[width=0.5\textwidth,height=\textheight,keepaspectratio]{res_final_phot_NH15.pdf}\\
\includegraphics[width=0.5\textwidth,height=\textheight,keepaspectratio]{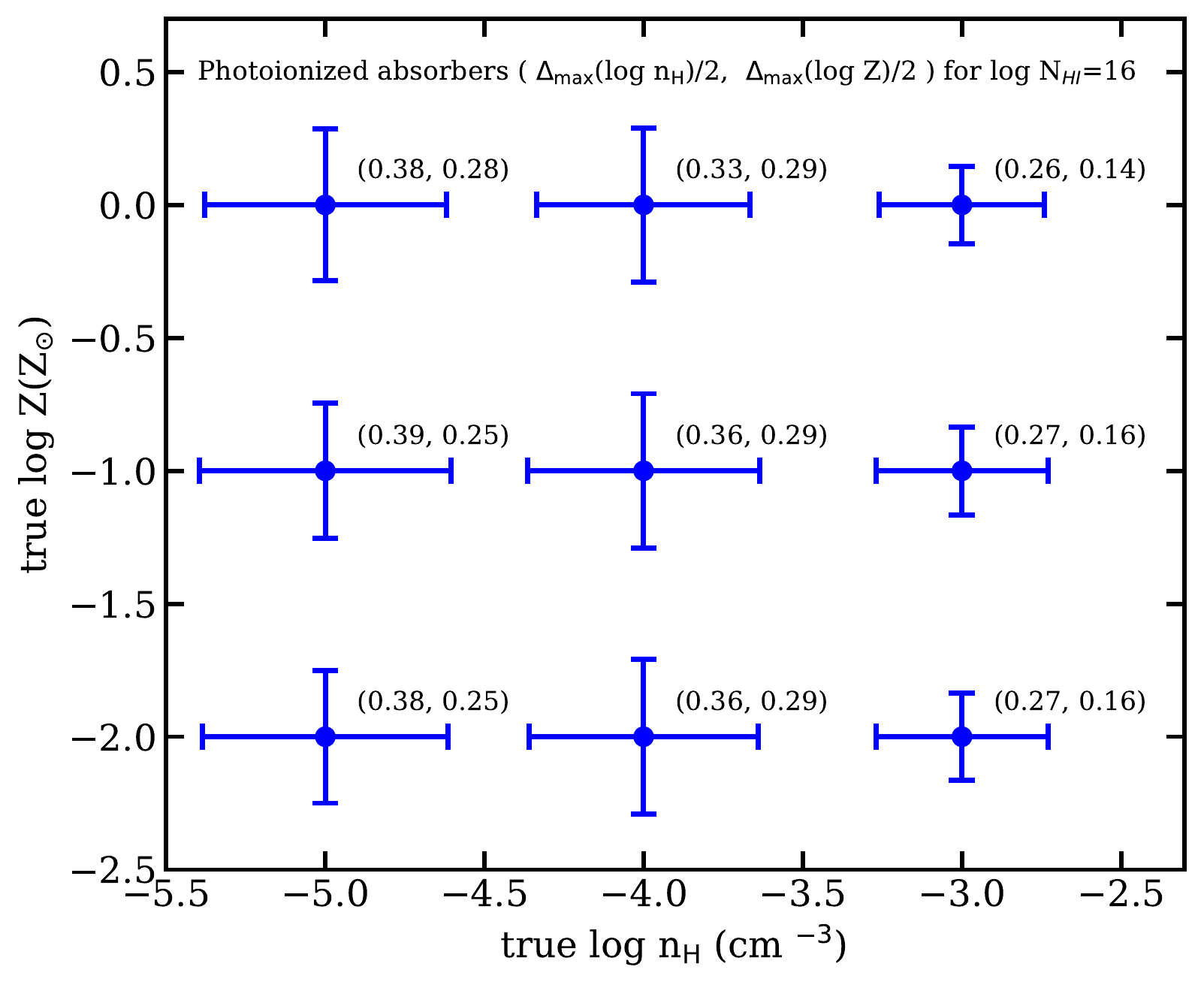} &
\includegraphics[width=0.5\textwidth,height=\textheight,keepaspectratio]{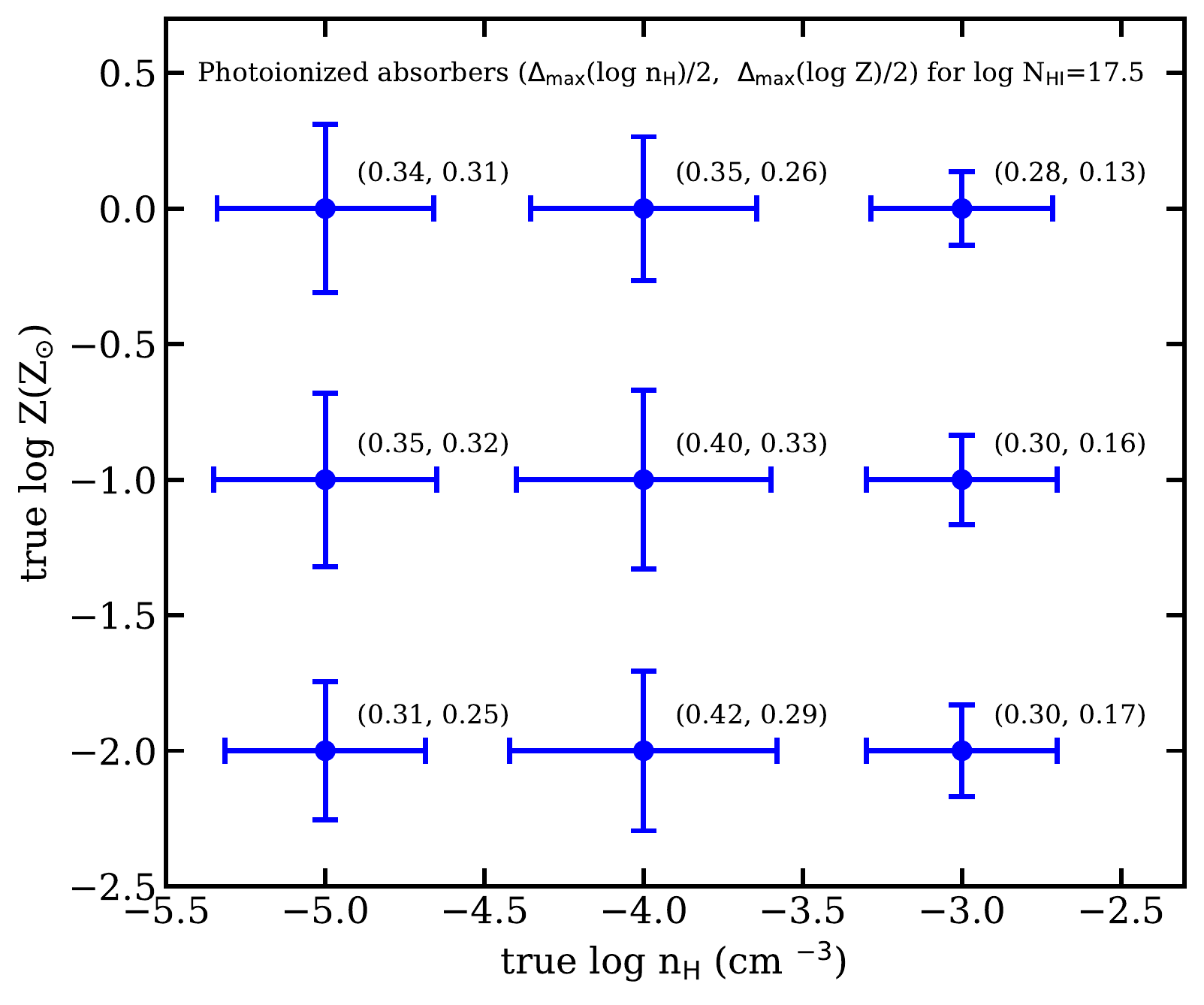} \\
\includegraphics[width=0.5\textwidth,height=\textheight,keepaspectratio]{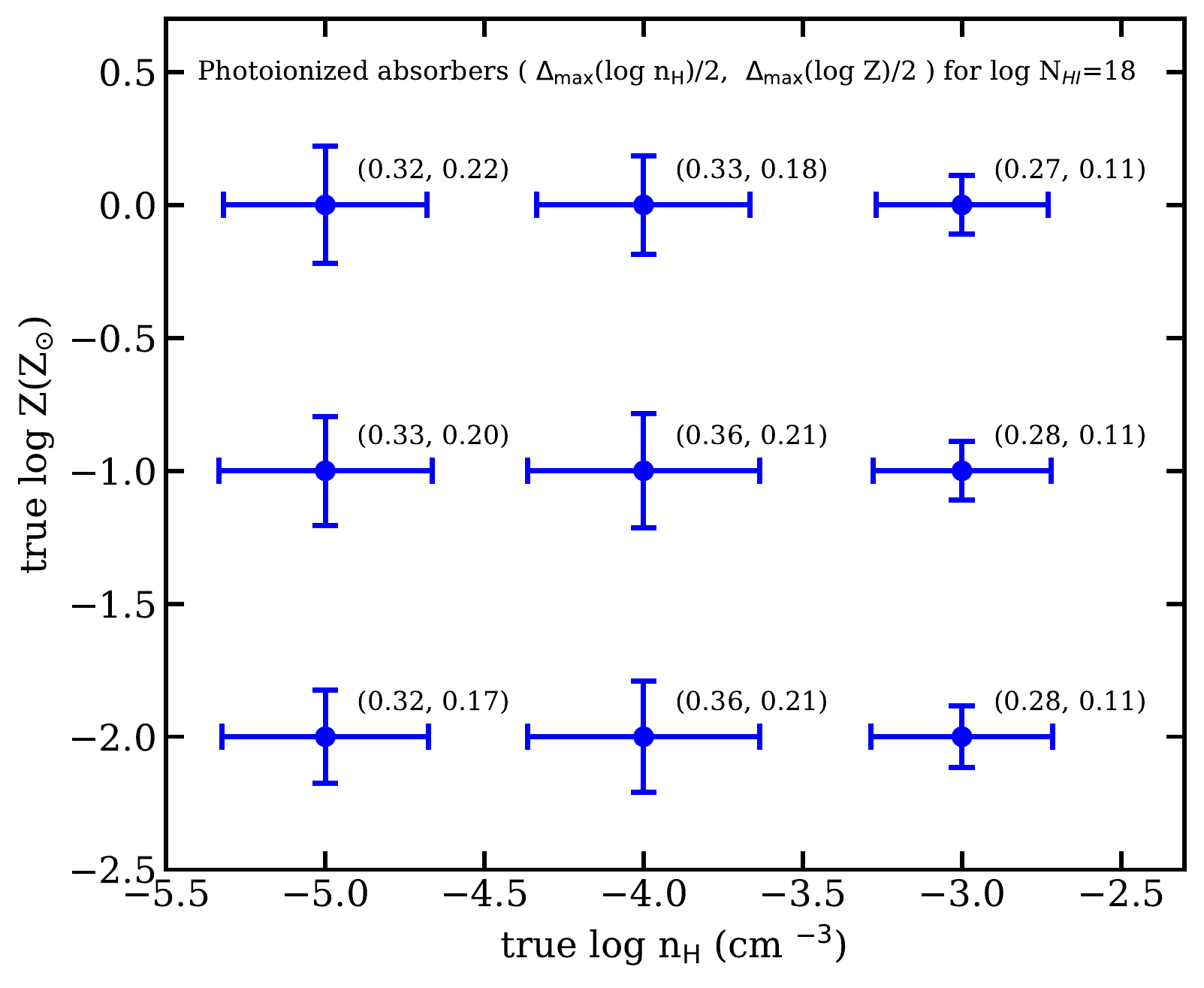} &
\includegraphics[width=0.5\textwidth,height=\textheight,keepaspectratio]{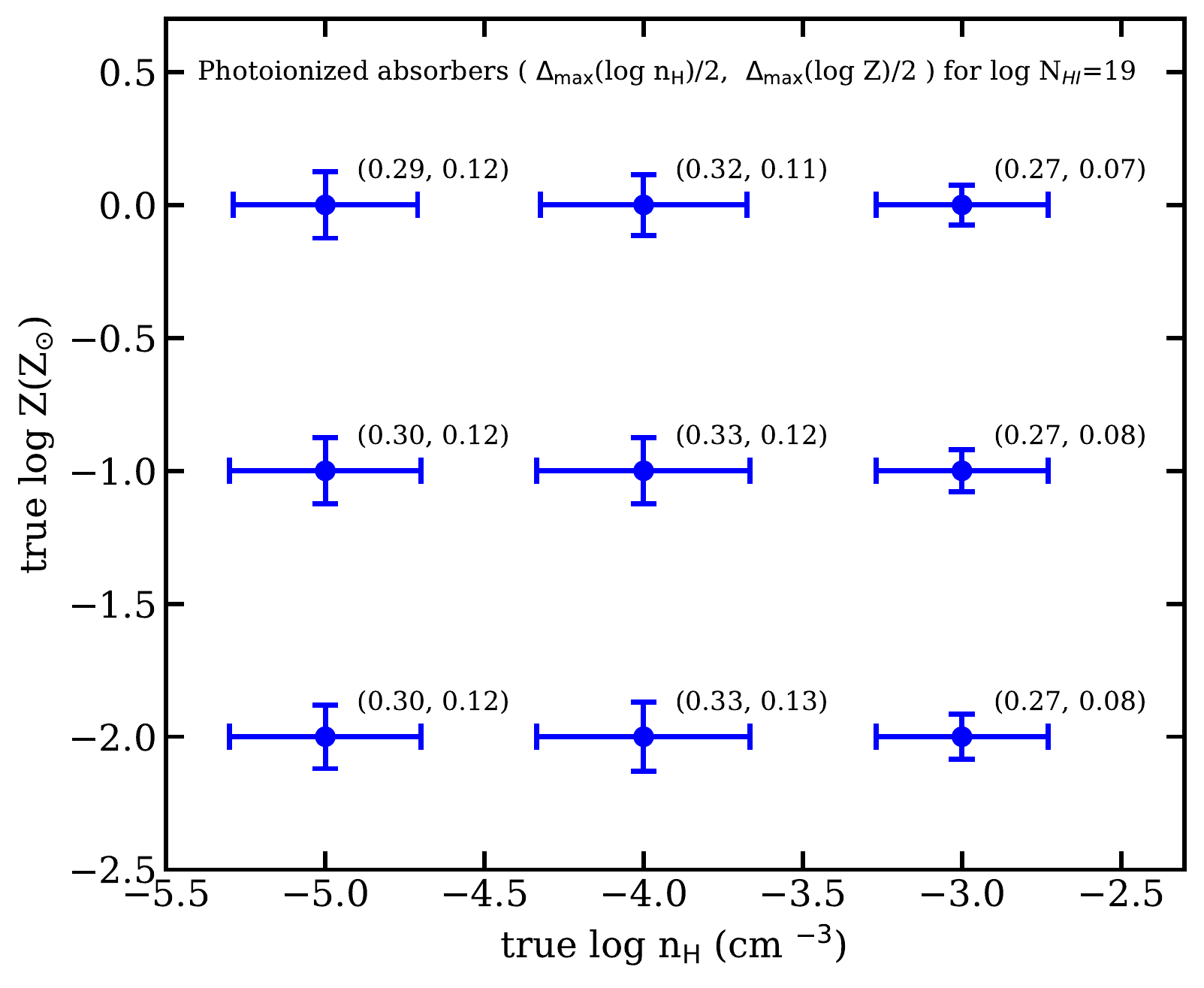}
\end{tabular}
\caption{ Results for a range of different H~{\sc i} column densities 
log$N_{\rm HI} = 10^{14}$ cm$^{-2}$ to log$N_{\rm HI} = 10^{19}$ cm$^{-2}$. Note that optical thickness of the cloud does not seem to have a significant effect on the results. Note that the result using a stopping column density of log$N_{\rm HI} = 10^{15}$ cm$^{-2}$ are shown in figure \ref{fig.res_photo_final}.
 }
\label{fig.FigB}
\end{figure*}
\end{sloppypar}
\label{lastpage}
\end{document}